%% file: main_JOE_CURRENT.tex
\documentclass[12pt,reqno]{amsart}
\pdfoutput=1
\input{header/packages}

\theoremstyle{plain}

\theoremstyle{definition}

\newtheorem{remark}{Remark}

\input{header/macros.tex}

\begin{document}
\pagenumbering{gobble}
\bigskip

\title{Hedonic Prices and Quality Adjusted Price Indices Powered by AI}
\author[Bajari et al.]
{P. Bajari$^*$, Z. Cen$^*$, V. Chernozhukov$^*$, M. Manukonda$^*$, S. Vijaykumar$^*$, J. Wang$^*$}\author[]{  R. Huerta, J. Li, L.Leng, G. Monokroussos, S. Wang}
\thanks{$^*$P. Bajari, Z. Cen, V. Chernozhukov, M. Manukonda, S. Vijaykumar, and J. Wang (in alphabetical order) are the principal authors who contributed to the current version of this paper. R. Huerta
and G. Monokrousos were also principal contributors to an earlier version of the paper. 
For helpful comments, we would like to thank the participants at the
2018 Federal Economic Statistics Advisory Committee meeting, the 2019 Allied Social Science meetings, the 2019 Federal Reserve Board conference on "Nontraditional Data, Machine Learning, and Natural Language Processing in Macroeconomics", the 2019 Brookings Conference "Can big data improve economic measurement?" and the 2021 CEMMAP conference "Measuring prices and welfare,"  and seminars at Berkeley, European Bank of Reconstruction and Development, MIT, UCL, and York.  We are grateful to Andrew Chesher, Greg Duncan, Kevin Fox, John Haltinwager, James Heckman, Daniel Miller, Mathew Shapiro, Bernhard Schölkopf, James Stock, and Weining Wang  for helpful comments during the various stages of this project. The paper was first publically circulated in 2021 as a CEMMAP Working Paper CWP04/21.
 } 
\date{March, 2019. This version: \today.}
\begin{abstract}
  We develop empirical models that efficiently process large amounts of unstructured product data (text, images, prices, quantities) to produce accurate hedonic price estimates and derived indices.  
  To achieve this, we generate abstract product attributes (or ``features'') from  descriptions and images using deep neural networks. 
  These attributes are then used to estimate the hedonic price function.  
  To demonstrate the effectiveness of this approach, we apply the models to Amazon's data for first-party apparel sales, and estimate hedonic prices. 
  The resulting models have a very high out-of-sample predictive accuracy, with $R^2$ ranging from $80\%$ to $90\%$. 
  Finally, we construct the AI-based hedonic Fisher price index, chained at the year-over-year frequency, and contrast it with the CPI and other electronic indices.  \\

  Key Words: Hedonic Prices, Price Index, Transformers, Deep Learning, Artificial Intelligence
  
  \end{abstract}

\maketitle

\newpage
	
\newpage
\pagenumbering{arabic}
\section{Introduction}
 Economists and policy-makers rely on price indices, such as the Consumer Price Index (CPI), to measure inflation, consumer welfare, and the cost of living. As a result, the methods used to construct price indices warrant considerable attention.
Two standard methods are the Laspeyres and Paasche indices, which measure changes in the cost of a standardized basket of products between two
periods, where the basket is chosen to represent aggregate demand in either the initial
period (Laspeyres) or the final period (Paasche). The two are often combined into
what is known as the Fisher Price Index (FPI).\footnote{These quantities bound other measures of inflation based on the expenditure function under certain assumptions; see \cite{diewert1998index} and also \cite{diewertfox2022substitution} for a state-of-art review.} 
 The FPI has been shown to accurately approximate the relative cost of living between two periods when the difference in prices is small.\footnote{The FPI captures the cost of living for a representative consumer with a quadratic utility or expenditure function, and provides a second-order approximation for any smooth utility or expenditure function under small price changes (see \citealt{diewert1976exact}). Price indices with this property are called superlative indices, with another prominent example being the Tornqvist index (see, e.g, \citealp{office_of_national_statistics_using_2020}). For both matched and hedonic indices, the Tornqvist indices we obtain are numerically very similar to the corresponding (matched or hedonic) Fisher indices.
}

A common problem with the aforementioned indices is product entry and exit, where the previous period's prices are not available for new products, and vice-versa.  Consequently, economists often produce so-called `matched indices': indices restricted to the set of products that are bought and sold in both periods. This introduces selection bias
because products exiting the marketplace may not resemble those that remain, especially when products turn over quickly.  To mitigate this bias, economists often use high-frequency chaining combined with compounding: for example, computing price indices at a monthly frequency and then compounding monthly inflation rates to get yearly rates.  This approach does mitigate the turnover problem if the rate of turnover is not high from month to month. Still, it can suffer from chain-drift bias\textemdash bias in estimated price levels due to geometric compounding of measurement errors.

Hedonic price models were introduced by \cite{court1939dynamics} and \cite{griliches1961hedonic}; these models postulate that the prices of differentiated products are determined by the market value of each product's underlying characteristics.
Court and Griliches suggested measuring inflation or deflation by modeling how hedonic prices change while holding product characteristics fixed. Importantly, one can compute hedonic prices for a basket of goods at any time point\textemdash despite entry and exit\textemdash because such prices are determined only by product characteristics. These prices can then be used in the Fisher price index and other price index formulas, giving rise to hedonic price indices. Hedonic price indices are regularly employed both in academic research and by statistical agencies (e.g.,  \citealp{moulton-hedonics,office_of_national_statistics_using_2020}).

The resulting hedonic price indices are sometimes called \emph{quality-adjusted} price indices because we implicitly fix the set of characteristics (``qualities") of a basket in a reference period and compute the ratio of costs of the basket in the comparison period and in the reference period. The ability to compute hedonic prices at any time point for any product allows us to address the entry/exit problem and also reduce chain-drift biases by making "long" or ``low-frequency" comparisons (year-to-year, for example).\footnote{Year-to-year chaining is recommended 
in the CPI manual \citep{CPImanual} to deal with chain drift. This approach was used, for example, in \cite{handbury} to construct non-hedonic indices using electronic data from Japan. See \citet{diewertfox2022substitution} for pertinent discussion and other methods used to address chain drift. One prominent class of alternative methods are multilateral indices\textemdash including the GEKS index\textemdash which average many matched indices with varying base periods. 
}  Success of this approach depends both on the ability of hedonic price models to approximate real-world prices and on our ability to accurately estimate the hedonic price function.  


Provided that hedonic models provide a good approximation of real world supply and demand, we can hope to model equilibrium prices by regressing observed prices on product characteristics, $X_j$. In traditional empirical hedonic models, the construction of $X_j$ is performed using human expertise, and statistical agencies perform the data collection through extensive field surveys and interviews. 

In this paper, we develop an alternative approach to building hedonic models. 
Our method uses electronic data and artificial intelligence to collect real time data and automatically generate product characteristics, $X_j$, which are used to accurately predict prices. 
The resulting approach is highly scalable and cost-effective compared to the traditional approach. Indeed, the question of how to leverage electronic records to improve the construction of price indices has received substantial attention from both academic researchers and statistical agencies (see e.g.~\citealp{groves2017innovations,ehrlich2019minding,jarmin2019evolving,lebow2023modernizing}). 
Our approach thus offers a valuable complement to traditional methods for measuring price-level changes.

\subsection{Deep Learning and Structured Sparsity} 

\begin{figure}[t]
\begin{center}
\includegraphics[width=4in]{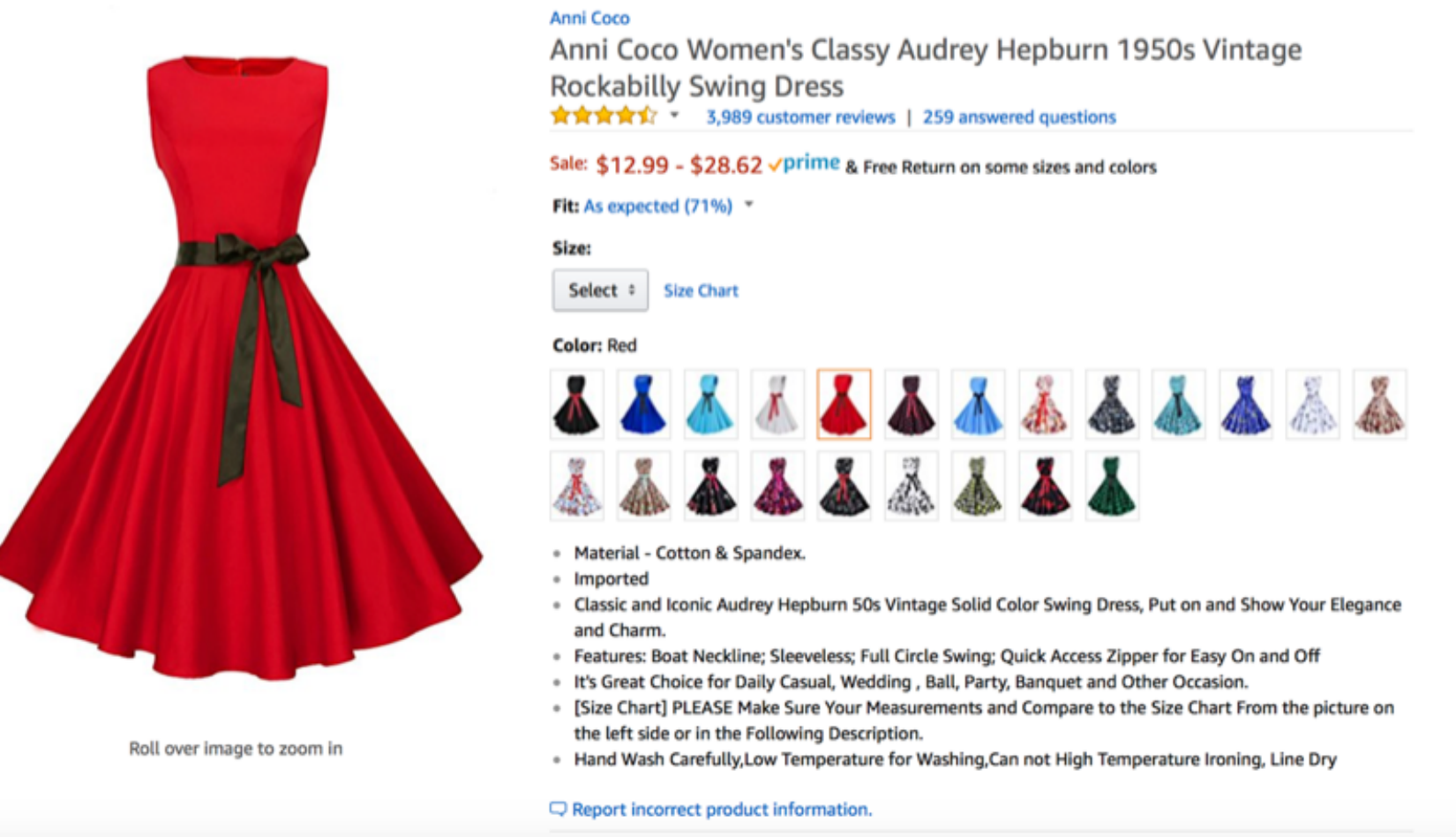}
\par\end{center}
\caption{An example of product characteristics for a product sold in the Amazon store.}
\label{fig:detail-page}
\end{figure}

Let us illustrate the problem we are solving. Figure \ref{fig:detail-page} shows the product characteristics visible to customers, including the product title, description, and images. Our goal is to represent this information using a numerical vector $X_j$ of moderately high dimension (in our case it has 2000 entries), which can be used to predict prices accurately. Moreover, the representation needs to be algorithmic and scalable.

The success of this approach depends on the existence of parsimonious structures behind images and text.  Traditionally, analysts relied on human experts to represent the key features of products by means of a low dimensional numerical representation, $X_j$.  Successful experts did produce low-dimensional representations for certain groups of products, which proved to be successful in building hedonic models (e.g., \citealp{pakes2003reconsideration} reports very high accuracy for predicting computer prices).  However, this approach does not scale well to many types of products and is prone to judgment biases. These issues raise important questions: when human experts do succeed, what is the underlying reason?;  can we replicate this success with artificial intelligence, and can these methods deliver scalable inference?

Arguably, humans can easily summarize the red dress depicted in Figure 1 and its accompanying text, even though the original representation of this information lives in an extremely high-dimensional space. Indeed, the image consists of nearly one million pixels (three layers of 640 x 480 pixels encoding the blue, red, and green color channels), while words belong to a dictionary whose dimension is in the tens of thousands and sentences live in much higher-dimensional space \citep{milton2013vocabulary}. However, we believe that information in images and sentences can be effectively represented in a much lower-dimensional space, a phenomenon we call ``structured sparsity." Human intelligence can exploit this structured sparsity to process information effectively\textemdash perhaps using the geometry of shapes in images, the relative simplicity of color schemes and shade patterns, the similarity of many words in the dictionary, and the context-specific meaning of words. The field of artificial intelligence (AI) developed neural networks to mimic human intelligence in many information-processing tasks.  These models do create parsimonious structures from high-dimensional inputs and often surpass human ability in such tasks.  Going forward, we will employ state-of-the-art solutions from AI to the problem of hedonic modeling.

In this work we extract relevant product attributes, or ``features,'' from text and images using deep learning models. These features, which are denoted $V_j$, take the place of traditional hedonic features tabulated by human experts.
We then use these features to estimate the hedonic price function.
To generate $V_j$, we convert text information about the product to numeric features (embeddings) using the BERT model of \citet{devlin2018bert}, a transformer-based large language model; our version of the model has been fine-tuned on Amazon's product descriptions and prices.
We similarly use a pre-trained ResNet50 model, a convolutional neural network used to understand images, to produce embeddings for product images \citep{he2016deep}. 
For context, these models were initially trained to comprehend text and images in tasks unrelated to predicting prices (e.g., image classification, or prediction of a missing word in a sentence). We then take the internal numeric representation generated by each model as an information-rich, parsimonious \emph{embedding} of the input.\footnote{This strategy follows a paradigm called ``transfer learning,'' which has proved a very successful way to use neural networks in new domains \citep{ng2016nuts}.}
With these embeddings, we then estimate the hedonic price function using a neural network. In particular, we design \emph{multi-task} networks, which predict a complete time-series of hedonic prices $(\hat H_{it})_{t \in T}$ for a given product $i$, where $t \in T$ covers all time periods in the study.\footnote{This requires training the network using data for all time periods. See Section \ref{sec:hedonic-indices} for a comparison to other models which do not share this constraint, and Appendix \ref{appendix:alternative-nn-models} for further discussion.}


\begin{figure}[ht]\label{fig:process}
\begin{center}
\includegraphics[width=6in,height=3in]{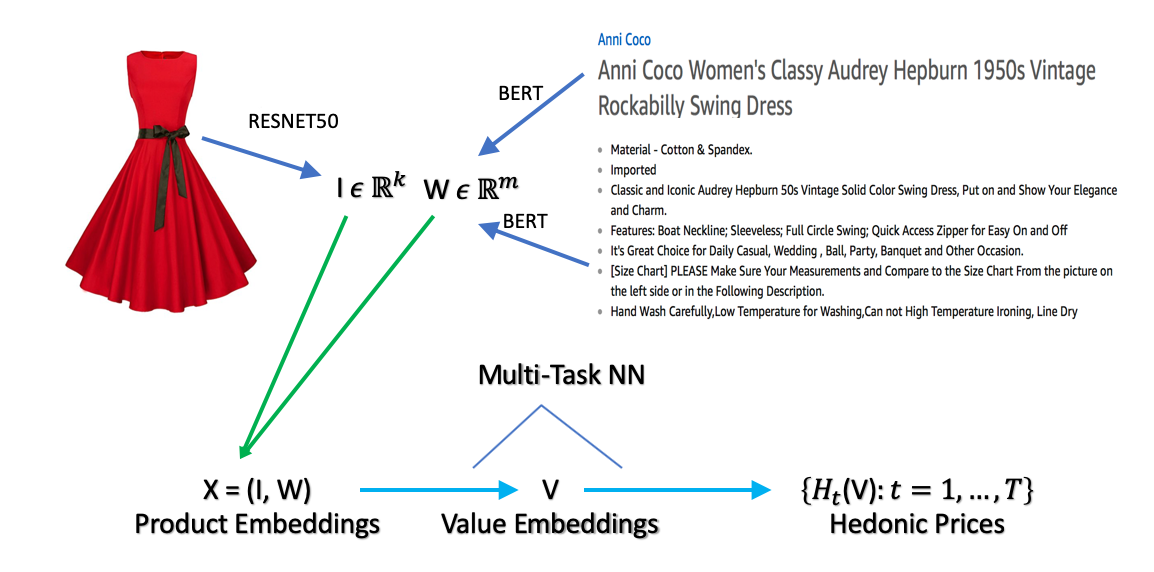}
\par\end{center}
\caption{\small Our method for generating hedonic price: The input consists of images and unstructured text data. The first step of
the process creates the moderately high-dimensional numerical
embeddings $I$ and $W$ for images and text data via state-of-the-art neural networks: ResNet-50 and BERT.  The second step takes input
$X = (I,W)$ and creates predictions for hedonic prices $H_t(X)$ using a multi-task neural network. Our multi-task model creates an intermediate lower dimensional embedding $V= V(X)$, called a \textit{value embedding}, and then predicts the final prices in all periods $\{H_t(V),  t=1,..., T\}$ using linear functional forms; this makes it easy to perform inference on the last step using hold-out data.}
\end{figure}

\subsection{Deriving Indices using Amazon Data} We apply these models to Amazon's data for apparel sales to estimate hedonic prices.\footnote{Our data cover ``first party''  sales: those in which Amazon first buys the product from a trusted seller and then sells it directly to the consumer.}  The resulting models have high predictive accuracy, with $R^2$ in the hold-out sample ranging from $80$ to $90\%$. Therefore, our approach can attribute up to $90\%$ of variation in price to variation in the product embeddings that encode the product attributes. We find this performance remarkable for two reasons:

\begin{enumerate} 
\item The production of hedonic prices is completely automatic and scalable, without relying on any human-based feature extraction.
\medskip
\item The performance suggests that hedonic price models provide a good, first-order approximation for real-world prices.
\end{enumerate}

We then proceed to construct Fisher hedonic price indices (FHPI) over 2013-2017, constructing monthly-chained, yearly-chained, and the geometrically combined FHPI (GFHPI).\footnote{The GFHPI gives equal geometric weight to yearly and monthly chained indices. This brings in some seasonality, allowing the GFHPI to reflect within-year price changes. This index should not be used over long horizons, however, due to chain drift.}\footnote{This approach was inspired by the GEKS index, which computes the geometric average over all chains; see e.g., \cite{diewertfox2022substitution} for a precise definition. The GEKS index reduces chain drift caused by measurement error in the product shares used to construct the monthly index. We are currently exploring the use of hedonic GEKS indices as follow-up work.}  We focus discussions on yearly chained FHPI and refer to it as the Fisher hedonic price index unless stated otherwise. We compare this index with
\begin{itemize}
\item the matched (repeated sale) Fisher index, 
\item the posted price Jevons index, which is the geometric mean of relative prices, chained at a daily frequency, 
\item the BLS Urban CPI index for apparel (CPI), constructed by the Bureau of Labor Statistics, 
\item the Adobe Digital Price Index (DPI) for apparel, constructed by \cite{goolsbee2018internet} using the Adobe Analytics data.
\end{itemize}

All indices suggest the apparel price level declines over this period. The annual rate of inflation estimated by FHPI is -.98\%, by the Jevons index -3.01\%, and by the matched Fisher index -3.12\%.  In comparison, the annual rate of inflation in apparel prices estimated by the Adobe DPI is -2.02\%, and by the CPI is -.31\%. 

The yearly chained FHPI, our preferred index, has the smallest discrepancy with the CPI. Part of the remaining difference with the CPI could be attributed to the limited ability of the  CPI to address quality change and substitution (as pointed out by \citealp{moulton2018measurement}, and others), to specifics of product categorization in the apparel segment, or to differences in the composition of baskets purchased by Amazon's consumers.  The difference may also reflect Amazon-specific cost improvements and supply-chain logistics. 

In comparison to the FHPI, neither the matched Fisher index nor Adobe’s DPI adjusts for change in the quality of goods sold. These indices potentially also suffer from chain drift\textemdash the systematic accumulation of errors due to frequent compounding.  Compared to the FHPI, the Jevons index does not incorporate quantity weighting, and it is also subject to chain drift.\footnote{Chain drift due to measurment error is less severe for the Jevons index\textemdash at least theoretically\textemdash because the index is computed using very many prices so that within-period errors are small. The Jevons index is a special case of the Tornqvist index with revenue shares for products set to be equal. This reduces another important source of chain drift due to measurement error in quantity weights, as pointed out by \cite{ivancic2011scanner}.} Both of these issues produce a large discrepancy between the Jevons index and the FHPI.  

\subsection{Contributions to the literature} We view our paper as an original contribution towards the modernization of hedonic price models and their application to  large-scale data. In this way, we also contribute to the empirical literature dedicated to hedonic price indices and their uses in measuring inflation. 

\subsubsection{AI-based hedonic models} {At a high level, our work stems from the observation that hedonic modeling is fundamentally a prediction problem: ``{how much demand will there be for characteristics $X_j$ at time $t$, based on the demand for similar products?}'' As such, it presents an excellent opportunity for AI to help us measure policy-relevant economic variables, such as the rate of inflation \citep{mullainathan2017machine}. This is especially significant given the massive amount of unstructured data available in the form of electronic records, which are not easily incorporated into conventional models.}

We are unaware of any prior work that develops large-scale hedonic price models from unstructured product text and image descriptions. In addition, our data are unique in that they cover the universe first party transactions in  Amazon's U.S.~stores from 2013 to 2019. From these models and data, we generate interesting findings: we demonstrate the power of hedonic models to characterize prices and document the decline in the quality-adjusted Fisher price index for apparel.  To the best of our knowledge, there are very few related studies in economics: An independent and contemporaneous work by \cite{zeng2020} develops a related approach to hedonic prices using scanner data, but uses random forest methods and low-dimensional data instead of neural network embeddings of unstructured text and images. An independent and contemporaneous work by \cite{han2021} explores the use of image embeddings to characterize typesetting fonts as products, and analyzes the effect of mergers on product differentiation via font producers' design choices. In the coming years, we expect to see see a much wider use of AI-based embeddings for text and images to power empirical research in economics.


\begin{table}
 \begin{center}
\begin{tabular}{l|cccc}
\hline\hline
Index  & CPI  & BPP  & {DPI}  &  FHPI  \\
\hline
Prices & Yes & Yes & Yes & Yes \\
Revenue Shares for Product Groups & Yes & Unknown & Yes & Yes \\
Quantities & No & No & Yes & Yes \\
Quality Adjustment & Yes & No & No & Yes\\ 
Long Chaining  & Yes & No & No &Yes\\
\hline\hline
\end{tabular}
\caption{ Some properties of the CPI, BPP, {DPI}, and FHPI}
\label{}
\end{center}

\end{table}

\subsubsection{Inflation, e-commerce, and electronic data} This paper contributes to a rapidly growing literature on the use of electronic data to measure inflation and other aggregate quantities. {This is motivated by the significant growth of e-commerce, along with evidence that online prices may fluctuate differently than their offline counterparts.}\label{subsec:online-indices-intro}

The MIT Billion Prices Project (BPP) constructs Jevons indices using web-scraped retail price data and directly-provided retailer data. The advantages of such data  include real-time availability at daily frequencies, low collection costs, large product counts, and uncensored price spells. Consequently, BPP indices constructed using modern real-time data can serve as useful benchmarks for official government statistics. For instance, \cite{cavallo2011distribution, cavallo2016billion} study several countries and establish that inflation measures constructed using such online data can differ substantially from official government statistics on consumer prices (the most extreme example being the case of Argentina, see \citealp{cavallo2013online}).  

A recognized limitation of this approach is that it does not incorporate quantity information. To this end, \cite{goolsbee2018internet} constructed matched T\"ornqvist indices, including the Digital Price Index (DPI), using Adobe Analytics data from e-commerce clients of Adobe, which notably include quantities as well as prices. This approach overcomes the quantity limitation of the BPP index and thus can account for substitution resulting from consumers' cost minimization. They find that the DPI for U.S. online transactions measures inflation to be substantially lower  than the official CPI's measurement for the categories they study in the period 2014-2019. This is analogous to our findings for apparel, albeit the discrepancy between our preferred index and the CPI is much smaller. The difference with our approach could possibly occur due to chain drift in the DPI due to monthly chaining.

 There is also a body of research on using scanner data to survey demand and to construct measurements of inflation; of particular note are \citet{handbury}, \citet{ivancic2011scanner}, \citet{leicester2014potential}, \citet{diewertfox2022substitution}, and \citet{office_of_national_statistics_using_2020}. We contribute to this development by providing AI-based hedonic prices that economists can employ alongside the approaches in these studies. {This provides a valuable complement, as scanner data are generally tilted towards food and beverages sold in grocery stores \citep{kaplan2017inflation}.} 
 
 In related work using online price data, \citet{cavallo2017are} finds that online prices are identical to their offline counterparts about 70\% of the time (on average across countries). \citet{gorodnichenko2017price}, \citet{cavallo2018scraped} and \citet{gorodnichenko2018price} find that online prices change more frequently than offline prices, thereby responding to competition more promptly. These papers also find that online prices exhibit stronger pass-through in response to nominal exchange-rate movements than prices found in official CPI data; such results have important implications for the price stickiness literature and the law of one price. These results also highlight the potential benefit of real-time, electronic data and derived price indices like ours.

 Since the circulation of an earlier version of this manuscript as a CEMMAP Working Paper \citep{bajari2019hedonic}, there has been substantial and very promising research that adapts our methods to other, publicly available data sources, with the ultimate aim of complimenting or improving national statistics. Of particular note are the working papers of \citet{cafarella2023using} and \citet{ehrlich2023quality}, both of which discuss the use of ML-based hedonic indices with point-of-sale scanner data.

\subsection{Organization of the Paper} We organize the rest of the paper as follows. Section 2 defines the hedonic price models and price indices. Section 3 discusses modeling and estimating the hedonic price functions via Neural Networks (NNs). Section 4 provides a non-technical description of the process of obtaining product embeddings via AI tools. Section 5 examines the empirical performance of the AI-based hedonic price functions and constructs the AI-based hedonic price indices.  Appendix A provides a technical description of BERT, a large language model used to generate embeddings for the product description. Appendix B gives a technical description of ResNet50, a tool used to generate embeddings of the product's image. Appendix C gives a technical description of various models we've considered and tested, of which the best-performing methods are described in the main text. 

\subsection{Notation} We use capital letters as $W$ as random vectors and $w$ the values they take; we use $\mathsf{W}$ to denote
matrices.  Functions are denoted by arrows $w \mapsto f(w)$ or simply $f$. Greek symbols denote parameter values, with the exception of $\epsilon$ which denotes the regression error (see below).

\section{Hedonic Prices and Hedonic Price Indices} \label{sec:hedonic-indices}

\subsection{The Hedonic Price Model}  

 We denote the product by its index $i$ and the time period (month) by $t$. An empirical hedonic model is a predictive model for the price given  product features:
\begin{equation}
P_{it}= H_{it}+ \epsilon_{it} = h_t(X_{i}) + \epsilon_{it}, \ \ \Ep [\epsilon_{it} \mid X_{i}] = 0,
\end{equation}
where $P_{it}$ is the price of product $i$ at time $t$, $X_{i}$ are the product features. The price function $x \mapsto h_t(x) $ can change from period to period, reflecting the fact that product features may be valued differently in different periods.  For our purposes, the advantage of these models is that they allow us to compare new goods to old rather directly; we simply compare the value consumers attach to the characteristics of the old good to those of the new.

We will use the data from time period $t$ to estimate the function $h_t$ using modern nonlinear regression methods, such as deep neural networks.
We contrast this approach with classical linear regression methods as well as other modern regression methods, such as a random forest.  The key component of our approach is  the generation of product features $X_{i}$ using neural network embeddings of text and image information about the product. Thus,  $X_{i}$ consists
of text embedding features $W_{i}$, constructed by converting the  title and product description into numeric vectors,
and image embedding features $I_{i}$, constructed by converting the product image into numeric vectors:
\begin{equation}
X_{i}= (W_{i}',I_{i}')'.
\end{equation}
These embedding features are generated respectively by applying the BERT and ResNet50 mappings detailed
in the next section.

There is a substantial body of economic research on hedonic price models. On the theory side, economists have developed a theory of supply and demand in terms of product characteristics \citep{lancaster1966,mcfadden1974measurement, rosen1974hedonic, gorman1980possible}; they have also established existence of the hedonic price function and characterized its behavior under various assumptions \citep{berry1995automobile,berry2004differentiated,ekeland2004identification,benkard2005hedonic,chiappori2010hedonic,chernozhukov2014single}; they have also developed the use of hedonic prices for measuring changes in consumer surplus and welfare \citep{bajari2005demand}.  On the empirical side, economists have estimated a variety of hedonic price models and linked them to the consumer's utility and their marginal willingness to pay for certain characteristics, see \cite{nesheim2006hedonic} and \cite{greenstone2017continuing}; they have also used hedonic theory to measure the value of non-tradable goods\textemdash for example, to measure the effects of air quality or hazardous waste cleanup on housing prices, (e.g.,~\citealp{chay2005does,stock1991nonparametric}), or to study equalizing differences in the labor market, (e.g.,~\citealp{brown1980equalizing}).
Our main use of hedonic prices is to estimate the rates of deflation (or inflation) for apparel products purchased at Amazon.com. This  follows prior work on using hedonic models to construct official price statistics (see e.g., \citealp{griliches1961hedonic,pakes2003reconsideration,moulton-hedonics,office_of_national_statistics_using_2020}), but with a major deviation: we use product features engineered using deep learning instead of using human experts to tabulate product characteristics, and we also estimate the hedonic price function using deep learning rather than using classical regression methods.

The literature typically specifies three building blocks of theoretical hedonic models: consumer utility functions defined over products' characteristics (rather than products themselves);  producer cost functions likewise defined over characteristics of the product; and an equilibrium assumption (or existence of the equilibrium is shown as a part of the analysis); see e.g.~\cite{pakes2003reconsideration} and \citet{rosen1974hedonic}.\footnote{Typically the utility and cost functions will also depend upon characteristics of the consumer and producer, respectively.} This determines prices and quantities given demand and input costs, and establishes the existence of the hedonic price function 
$$
(x, u) \mapsto H^\star ( x, u)
$$
as a function of product attributes $(x,u)$. Here, both $x$ and $u$ are observed by the consumer, while only $x$ (and not $u$) is observed by the modeler. 

Price functions give us information about customer preferences.  For example, when the customer's utility is given by:
$$
V(x,u, p,m) = V_0(x,u) + m - p
$$
where $p$ is the price of the product to be paid by the customer and $m$ is their income,  the first order conditions 
for the utility maximization problem $\max_{(x,u)} V_0(x,u) + m - H(x,u)$ is given by:
$$
\partial_{x_k} V_0(X,U)  = \partial_{x_k} H^\star ( X, U),
$$
where $\partial_{x_k} = \partial/\partial x_k$, where $x_k$ refers to the $k$-th component of the vector $x$ (for continuously varying attributes). Therefore, under suitable assumptions, a standard argument for identification of the average derivative of a structural function gives $$
\Ep [\partial_{x_k} H^\star ( X_j, U_j) | X_j]  =   \partial_{x_k}   h_t(X_j).\footnote{This holds, for example, under independence of observable and unobservable characteristics at equilibrium, see Appendix \ref{appendix:identification}.}
$$
In other words, the average marginal willingness to pay for a given characteristic is equal to the average derivative of the hedonic price map, and is identified by the derivative of the hedonic regression function. We remark here that the expectation is taken over the random variables' distribution at equilibrium; we can not say anything about features of the hedonic price map away from equilibrium.

Given a parametric consumer utility function, preference parameters can be recovered from first-order conditions provided that either (i) $ (x,u) \mapsto H^\star(x,u)$ is additively separable in  $(x,u)$, so that $\partial_{x_k} H^\star ( x, u) = \partial_{x_k}   h_t(x)$, does not
depend on $u$, or (ii) we can identify $H^\star$ and the unobservable $U$ by other means (for example, by making quantile or multivariate-quantile type assumptions on the way $U$ appears in $H^\star$).  For example, if the utility  is Cobb-Douglas over observed characteristics,
\(V_0(x,u, p) = \sum_{k=1}^K  \alpha_k \log (x^\star_{jk}) +  \beta g(u) - p,\)
then under additive separability, 
\(H^\star ( X, U) =  H^\star_0 ( X) + U\),
we recover 
\(\alpha_k = \partial_{x_k} h_t(X_j) X_{jk}\) for a consumer who has purchased product $j$. The distributions of consumer taste parameters $\alpha$ can be recovered under this type of modeling approach; see \cite{bajari2005demand} for further relevant discussion.  Here we have a different goal: we use hedonic prices to construct indices and thereby measure price level changes, following accepted practice in applied economics and in the work of statistical agencies (e.g., \citealp{moulton-hedonics,pakes2003reconsideration, office_of_national_statistics_using_2020}). 

\subsection{Price Indices: Hedonic  vs Matched} 

We focus on hedonic price indices and compare them to matched (repeated sale) price indices. The matched price index tracks changes in the price of a basket of products sold in both the base period and in subsequent time periods.  While the matched price method is subject to selection bias (due to product entry and exit), it ensures the index tracks goods from a common pool of products.  A major shortcoming of this method is that the common pool of products across time can be small and non-representative; this will be apparent in our data.

The hedonic price index replaces transaction prices with predicted values using a rich set of product characteristics (obtained using a combination of AI and machine learning methods). In principle, the hedonic approach captures changes over time in the value consumers assign to product attributes.\footnote{We refer the reader to \citet{aizcorbe2014practical} for further discussion of hedonic price indices and their application.} Hedonic approaches are especially helpful for predicting the prices of new goods and dealing with the entry/exit selection bias when product prices are undefined.  This is especially relevant in our case, where we observe a very high turnover of products. 

We consider three broad types of price index:
\begin{itemize}
\item The Laspeyres (L) type, which uses base period quantities for weighting the prices;
\item The Paasche (P) type, which uses current period quantities for weighting the prices;
\item The Fisher (F) type, which is the geometric mean of the L- and P-type indices.
\end{itemize}

One defines the L- and P-type matched indices as measures of the total rate of price change 
of a basket of matching products  from the current period
$t$ with a previous period $t-\ell$:
$$
R_{t, \ell}^{P,M}=\frac{{\textstyle \sum_{i\in\mC_{t} \cap \mC_{t-\ell}}P_{it}Q_{it}}}{{\textstyle \sum_{i\in \mC_{t} \cap \mC_{t-\ell}}P_{j(t-\ell)}Q_{it}}}; \ \ R_{t,\ell}^{L,M}=\frac{{\textstyle \sum_{i\in \mC_{t} \cap \mC_{t-\ell}}P_{it}Q_{i(t-\ell)}}}{{\textstyle \sum_{i\in \mC_{t} \cap \mC_{t-\ell}}P_{i(t-\ell)}Q_{i(t-\ell)}}};
$$
and the F-type index takes the form $$R_{t,\ell}^{F,M}=\sqrt{R_{t,\ell}^{P,M}\cdot R_{t,\ell}^{L,M}},$$ where 
$Q_{it}$ is the quantity of the product $i$ sold in month $t$,  $P_{it}$ is the average sales price for product 
$i$ at time $t$, $\mC_{t}$ is the set of all products with transactions at time $t$, $\mC_{t} \cap \mC_{t-\ell}$ is the match set, the set of all products with transactions both at time $t$ and at time $t-\ell$.

In a standard, representative consumer model, 
matched indices should obey the order restriction
$R^{L,M} \geq R^{P,M}.$ Even when aggregate quantities are not described by a representative consumer model, 
the relation often holds empirically. Similar intuition applies 
to hedonic indices. The two indices
may be combined using Fisher's ideal index, which is a superlative index: it measures the exact cost of living when the utility
function is quadratic and provides a second-order approximation to the cost of living at the given prices
when the utility function is smooth \citep{diewert1976exact}.

We define the L, P, and F-type hedonic indices similarly, as measures of the total rate of hedonic price change 
of a basket of product attributes from the current period
$t$ with the previous period $t-\ell$:
$$
R_{t,\ell}^{P,H}=\frac{{\textstyle \sum_{i\in \mC_{t}}H_{it}Q_{it}}}{{\textstyle \sum_{i\in \mC_{t}}H_{i(t-\ell)}Q_{it}}}; \ \ R_{t,\ell}^{L,H}=\frac{{\textstyle \sum_{i\in \mC_{t-\ell}}H_{it}Q_{i(t-\ell)}}}{{\textstyle \sum_{i\in \mC_{t-\ell}}H_{i(t-\ell)}Q_{i(t-\ell)}}}; \ \ R_{t}^{F,H}=\sqrt{R_{t}^{P,H}\cdot R_{t}^{L,H}}.\footnote{When hedonic prices follow a linear model $H_{it} = \theta_t' V_i$, where $V_i$ are product attributes, we have $\sum_{i\in\mC_t} H_{is}Q_{it} = \theta'(\sum_{i\in\mC_t} V_i Q_{it})$. Thus, the index may also be viewed as measuring the relative price of a quantity-weighted basket of product attributes.}$$  We note that the index $P$ is defined over sets of products $\mC_{t}$ and 
the $L$ index is defined over the set of products $\mC_{t-\ell}$, which are supersets of the matching set $\mC_{t} \cap \mC_{t-\ell}$. 

The above indices exclusively use hedonic prices $H_{it}$, even when true prices $P_{it}$ are available; this is known as ``full imputation.'' Alternative approaches, particularly ``single imputation'' and ``double imputation,'' make use of observed prices $P_{it}$ for products in the matching set $\mC_{t} \cap \mC_{t-\ell}$ \citep[Sec. 3.2]{aizcorbe2014practical}. In subsequent work on ML-based hedonic indices, \citet{ehrlich2023quality} finds that full imputation tends to reduce chain drift, as estimated hedonic prices are less volatile than true prices. For our preferred index, both strategies produce nearly indistinguishable results.

For an arbitrary index $R_{t,\ell}^{\bullet,\bullet}$, for positive integers $t,\,\ell$, we measure the price changes up to time $t \ge t_0$ as follows. Let $m = \lfloor (t-t_0)/\ell \rfloor$ denote the largest integer no greater than $(t-t_0)/\ell$, and write $t-t_0 = m\ell + r$. We construct a chained index by taking the product
\[R^{\bullet, \bullet, C}_{t,\ell} = R^{\bullet, \bullet}_{t,r} \prod_{\bar m=1}^m R^{\bullet, \bullet}_{t_0 + \bar m\ell, \ell} \quad { } \text{ where }  \quad R^{\bullet, \bullet, C}_{t,0} =1.\footnote{For long-chained indices with $\ell > 1$, this approach compares a given period's prices to the most recent ``pivot'' $\{t_0, t_0+\ell, t_0+2\ell, \cdots\}$. This is a standard practice, see e.g.~\citet[Appendix 4]{statistics2022explanation}. Resulting indices are nearly identical to those given by alternative long-chaining methods.}\]



For the hedonic index we shall use month-over-month chaining with $\ell=1$ and  year-over-year chaining with $\ell=12$, getting
two types of indices:
$$
R^{F, H, C}_{t, 1}  \text{ and }   R^{F, H, C}_{t, 12},
$$
where the first index captures month-over-month changes in prices, especially for non-seasonal apparel, and the second index 
measures year-over-year changes in prices for products sold in the same month as $t_0$, including seasonal apparel. The second index is also less susceptible to the chain-drift problem that arises from an accumulation of errors due to repeated compounding.  For this reason, we choose the yearly-chained index to be our preferred index, and refer to it as the Fisher Hedonic Price Index (FHPI).  To capture seasonality while mitigating chain drift, we also consider the geometric mean of the monthly-chained and yearly-chained index (GFHPI): 
$
R^{GF, H}_t = \sqrt{R^{F, H, C}_{t, 1} R^{F, H, C}_{t, 12}}.$

\section{Price Prediction and Inference with Deep Neural Networks}
\label{sec:model}

 \subsection{The Multi-Price Prediction Network} \label{subsec:network-overview}
 Our model takes in high-dimensional text and image features as inputs, converts
 them into a lower-dimensional vector of value embeddings using state-of-the-art deep learning methods, and then outputs
 simultaneous predictions of price in all periods. 
 
Our general nonlinear regression model is a composition of several nonlinear, vector-valued functions, called \emph{layers}. It takes the form
  \begin{equation}\label{NNP}
Z_{i} = 
\left [ \begin{array}{l}
\mathrm{Text}_i \\
\mathrm{Image}_i
\end{array} \right]
 \overset{e}  \longmapsto  X_i  \overset{g_1}  \longmapsto  E^{(1)}_i \cdots\,  \overset{g_m}  \longmapsto  E^{(m)}_i =: V_i 
  \overset{\theta'} \longmapsto   \{  H_{it} \}_{t=1}^T := \{\theta_t' {V_i}\}_{t=1}^T. \\
\end{equation}
Here $Z_i$ is the original input, which lies in a very high-dimensional space. $Z_i$ is non-linearly mapped, via the embedding layer $e$, to an embedding vector $X_i$  which is of moderately high dimension (up to 5120 dimensions). This embedding is again non-linearly mapped to a lower dimension vector $E^{(1)}_i$ by the first hidden layer $g_1$, and so on, until the final hidden layer $g_m$. The output of the final hidden layer $g_m$, given by $V_i \coloneqq E^{(m)}_i$, is then \emph{linearly} mapped
to the final output, consisting of hedonic price $H_{it}$ for product $i$ in all time periods $t=1,..., T$.

The output of the final hidden layer, $V_i = E_i^{(m)}$, is called the \textit{value embedding} in our context. It is a moderately high-dimensional summary of the product (up to 512 dimensions, with 256 in our primary model). It is derived from product attributes, and directly determines the predicted hedonic price of the product.  Note that the embeddings $V_i$ do not depend on time and thus represent the intrinsic, potentially valuable attributes of the product. However, the predicted price does depend on time $t$ via the coefficient $\theta_t$, reflecting the fact that these intrinsic attributes are valued differently across time. 

The network mapping (\ref{NNP}) makes use of the repeated composition of nonlinear mappings of the form
\begin{equation}
g_\ell:  v \longmapsto  \{E_{k,\ell}(v)\}_{k=1}^{K_\ell} := \{\sigma_{k,\ell} (v' \alpha_{k,\ell})\}_{k=1}^{K_\ell},
\end{equation}
 where the $E_{k,\ell}$'s are called neurons, and $\sigma_{k,\ell}$ is the activation function that can vary with the layer $\ell$
 and can vary with $k$, from one neuron to another.\footnote{The standard architecture has an activation function
 that does not vary with $k$, but some architectures such as ResNet50 (discussed in Section \ref{sec:embeddings}) can be viewed as having
 an activation function depending on $k$, with some neurons linearly activated and some non-linearly activated.}  Standard examples include 
 the sigmoid function: $\sigma(v)=1/(1+e^{-v})$, the rectified linear unit function (ReLU),  $\sigma(v) = \max(0,v)$, or the linear function $\sigma(v)=v$. Individual neurons' activations can be linear or non-linear. The use of a non-linear activation function has been shown to be an extremely powerful tool for generating flexible functional forms, both yielding successful approximations in a wide range of empirical problems and backed by approximation theory. Good approximations can be achieved by considering sufficiently many neurons and layers (e.g., \citealt{chen1999improved,yarotsky2017error,kidger2020universal}).

Our empirical model uses up to $m=3$ hidden layers, not counting the input. The dimensions of each layer are described in Appendix \ref{appendix:alternative-nn-models}. The first layer is a sophisticated embedding of the input, trained using auxiliary text and image comprehension tasks using a separate dataset; further details are provided in Section \ref{sec:embeddings}.

The model can be trained by minimizing the loss function
\begin{equation}\label{eq:loss}
\min_{\eta \in \mathcal{N}, \{\theta_t\}_{t=1}^T}  \sum_t \sum_{i } (  P_{it} - \theta_t' {V_i} (\eta) )^2 Q_{it},
\end{equation}
where $\eta= (g_1,..., g_m)$ denotes all of the parameters of the mapping $X_i \mapsto {V_i} \eqqcolon V_i(\eta).$  Here we are weighting by the quantity $Q_{it}$. Regularization can be used to limit
 fluctuations of predicted prices across time. This is done by adding a penalty 
to the loss function:
\begin{equation}\label{eq:penalty}
\lambda \sum_i \sum_{t=1}^{T-1} | \theta_{t+1}' {V_i} (\eta) - \theta_{t}' {V_i} (\eta)|,
\end{equation}
where the penalty level $\lambda$ is chosen to yield good performance in the validation sample. We discuss the sample splitting procedure below in our description of model training.

Next, we give an overview of how the initial embedding is generated. A multilingual BERT model is used to convert text information into a sub-vector $W_i$ of $E^{(1)}_i$, and likewise a ResNet50 model is used to convert
images into another sub-vector $I_i$ of $E^{(1)}_i$ (both models are publicly available; see Section \ref{sec:embeddings} below). These models are trained on auxiliary prediction tasks with auxiliary output $A_{T_i}$
for text and $A_{I_i}$ for image, which can be illustrated diagrammatically as:
\begin{equation}
Z_{i} =  
\left [ \begin{array}{l}
\mathrm{Text}_i \\
\mathrm{Image}_i
\end{array} \right]
  \overset{e}  \longmapsto X_i:= \!\!\!\!\! \begin{array}{c} A_{T_i} \\ {\Big \uparrow} \\ \begin{bmatrix} W_i \\  I_{i}\end{bmatrix} \\ {\Big \downarrow} \\ A_{I_i}  \end{array}  \mapsto E^{(1)}_i \mapsto \cdots \mapsto  E^{(m)}_i := V_i
  \overset{\theta'}\longmapsto  \{ H_{it} \}_{t=1}^T. \\
\end{equation}
The text and image embeddings $W_i$ and $I_i$, which form $X_i$, are obtained by a procedure known as transfer learning. They are extracted from separate neural networks which were trained to map text or images to auxiliary outputs $A_{Ti}$ or $A_{Ii}$; in training, these auxiliary outputs were scored on natural language processing tasks and image classification
tasks, respectively. This first step was performed by other researchers and used data unrelated to prices\textemdash its aim was to produce versatile, high-quality embeddings $W_i$ and $I_i$ for general text and image data, as we elaborate upon in Section \ref{sec:embeddings}. Finally, we further fine-tuned parameters of the mapping that generates $W_i$ for price prediction tasks, yielding some improvements.\footnote{We did not attempt to fine-tune image embeddings $I_i$.} 

The estimates are computed using a sophisticated stochastic gradient descent
  algorithm. Such sophistication is needed because the optimization
  is generally not convex, making computation difficult. For the price prediction
  task, we used the Adam algorithm \citep{kingma2014adam}. At a high level, training involves randomly splitting the set of products into three subsets: training (60\%), testing (20\%) and validation (20\%).  This split occurs exclusively across products $i$, while preserving each product's time-series of prices and quantities.  The training and validation sets are used by the Adam algorithm to minimize the penalized loss given in Equations \eqref{eq:loss} and \eqref{eq:penalty}, while the testing set is used to measure predictive performance in Section \ref{sec:empirics}. The overall process has many tuning parameters; in practice, we 
  chose them by cross-validation. The most important choices concerned
  the number of neurons and the number of neuron layers.


\begin{figure}
\begin{center}
\begin{minipage}{.6\textwidth}
\begin{tikzpicture}[shorten >=1pt,->,draw=black!50, node distance=2em]
    \tikzstyle{neuron}=[rectangle,fill=black!25,minimum size=20pt,inner sep=0pt]

    \tikzstyle{layer}=[rectangle,fill=black!25,minimum size=20pt,inner sep=0pt]

    \tikzstyle{input neuron}=[neuron, fill=green!60];
    \tikzstyle{output neuron}=[neuron, fill=magenta!60];
    \tikzstyle{hidden neuron}=[neuron, fill=blue!60];
    \tikzstyle{annot} = [text width=6em, text centered];

    \foreach \y/\ypos in {1/2,2/3,3/4} {
        \node[input neuron] (I-\y) at (\ypos,-.5) {$Z_{\y}$};
        \node[output neuron] (oh-\y) at (\ypos cm, 4.75) {$O_{\y}$};
    }

    \foreach \y in {1,...,5} {
        \foreach \z / \zpos in {1/1,2/2.25, 3/3.5} {
            \node[hidden neuron] (h-\y-\z) at (\y cm, \zpos) {$E_{\y\z}$};
        };  
    
        \foreach \yy in {1,...,3} {
            \path (I-\yy) edge (h-\y-1);
            \path (h-\y-3) edge (oh-\yy);
        }
        
        \foreach \yy in {1,...,\y}{
            \path (h-\y-1) edge (h-\yy-2);
            \path (h-\yy-1) edge (h-\y-2);
            \path (h-\y-2) edge (h-\yy-3);
            \path (h-\yy-2) edge (h-\y-3);
        };
        
    };

    
    \node[annot] (hidden1) at (-1, 2.25) {Hidden Layers};
    \node[annot] (input) at (-1,-.5) {Inputs};
    \node[annot] (output) at (-1, 4.75) {Output};
\end{tikzpicture}
\end{minipage}
\par\end{center}
\caption{Standard architecture of a Deep Neural Network. In the hedonic price prediction network, the penultimate layer is interpreted
as an embedding of the product's hedonic value and the output layer contains predicted hedonic prices in all time periods. In comparison, the networks used for text and image processing have very high-dimensional inputs and outputs, with intermediate 
hidden layers composed of neural sub-networks. The dense embeddings typically result from taking the last hidden layer of the network.}\label{fig:NN}
\end{figure}

We depict the process conceptually in Figure \ref{fig:NN}, where we have a regression problem, and the network
  illustrates the process of taking raw regressors and transforming them
  into outputs, the predicted values.  In the first row we see the inputs, and in the second row we see the first layer of neurons. 
  The neurons are connected to the inputs, and the connections represent
   coefficients.  Finally, the last layers of neurons are combined  to produce a vector of outputs.   
  The coefficients $\theta_m$ are shown by the connections between
  the last hidden layer of neurons and the multivariate output.  Networks with vector
  outputs are called multi-task networks.

Prediction methods based on neural networks with many layers
of neurons are called deep learning methods.  Neural networks recently emerged as a powerful and all-purpose method for a wide range of problems\textemdash ranging from prediction and classification analysis to natural language processing tasks.  Using many neurons and multiple layers gives rise to deeper networks
  that are very flexible and can approximate the best prediction or classification rules
  very well in settings where data are plentiful. In Section 4 and Appendix, we overview the ideas and details of the neural networks for dealing with text and images.
  
   \subsection{Assessing Statistical Significance and Confidence Intervals} 
In many settings, researchers may wish to construct standard errors for predicted prices or for model coefficients.
To this end, the last layer of the neural network provides us with what we interpret as ``value embeddings,"
$$V_{it} =  (V_{1t},..., V_{pt})',$$
where we have chosen $p=256$ in our primary specification.  We condition on the training and validation data so that $V_i$ is considered fixed for all products $i$. Then we use the hold-out data to estimate the following linear regression model:
$$
P_{it} = V_{it}'\theta_t+  \nu_{it}, \quad \theta_t = (\theta_{1t},..., \theta_{pt})'.
$$
This is a low-dimensional linear regression model for which we can apply standard inference tools.
Applying linear regression to the test data, we obtain an estimate $\hat \theta_t$ and an estimated hedonic price 
\begin{equation}\hat H_{it} = V_{it}'\hat \theta_t.
\end{equation}
These additional estimates $\hat H_{it}$ and $\hat\theta_t$ will generally differ from $H_{it}$ and $\theta_{it}$ estimated by our primary model. Their main purpose is to allow the construction of predictive confidence intervals.\footnote{When the training and test datasets are many orders of magnitude larger than the dimension of $V_{it}$\textemdash as in our setting\textemdash the differences are generally small. In some settings, the updated predictions $\hat H_{it}$ may be preferable: they are not biased by the regularization in \eqref{eq:penalty}, and they lie at the center of the resulting confidence intervals.}

 For example, the statistical significance of features can be assessed by testing whether the regression coefficients $\theta_t$ is equal to zero, using $p$-values
and adjusting for multiplicity using the Bonferroni approach (or other standard approaches such as step-down testing methods or methods
that aim to control the false discovery rate).  We may also construct confidence intervals for individual coefficients, as well as level $1-\alpha$ confidence
intervals for the predicted hedonic price:
\begin{equation}
[L_{it}, U_{it}]= [\widehat H_{it}  \pm \Phi^{-1}(1- \alpha/2) \mathrm{SE}_{it}], \quad \mathrm{SE}_{it}= \sqrt{ V_{it}' \widehat{\mathrm{Cov}}(\hat \theta_t) V_{it}}.
\end{equation}

\begin{remark} The advantage of this approach is its simplicity, while the disadvantage is that it does not account for uncertainty in estimating
the value embeddings themselves (indeed, we consider them to be frozen conditional on the training and validation samples). Following \cite{genericML}, one  way to account for this variability is to consider multiple random splits, $s=1,...,S$, of the data into test, training and validation subsets (stratified by month). Different splits would result in different value embeddings
$V^s_{it}$, coefficient estimates $\hat \theta_t^s$,  predicted hedonic prices, and covariance estimates  $\widehat{\mathrm{Cov}}(\hat \theta_t)^s$,  as well as confidence intervals $[U^s_{it}, L^s_{it}]$. Then we can aggregate 
the estimates and confidence intervals as follows:
\begin{equation}
\widetilde H_{it}= \mathrm{median} \left (  (\hat H^s_{it} )_{s=1}^S \right),  \quad  \widetilde {CI}_{it} = 
\left [\mathrm{median} \left (  (\hat L^s_{it} )_{s=1}^S \right), \mathrm{median} \left (  (\hat U^s_{it} )_{s=1}^S \right) \right].
\end{equation}
The adjusted nominal level for the confidence interval for $V_{it}'\theta_s$ is $1-\alpha/2$.   Similarly, for judging the statistical significance of particular coefficients (or other functionals), we can consider multiple $p$ values $(P^{s})_{s=1}^S$ and aggregate them by taking the median
p-value and comparing this p-value to the adjusted nominal level $\alpha/2$.  This approach exploits the fact that the median
of arbitrarily correlated variables whose marginal distributions are standard uniform is stochastically dominated by the variable $2\cdot \mathrm{Uniform}(0,1)$; the resulting $p$-values are asymptotically uniformly distributed. We do not pursue this approach in the present paper as it requires training $S$ models instead of just one model. 

\end{remark}

\section{Image and Text Embeddings Via Deep Learning}
\label{sec:embeddings}
Typically, customers view products as depicted in Figure 1 of the introduction, where details such as product title, description, and images are presented. Our task is to convert these product characteristics into numerical vectors, which can be used for estimating hedonic prices, as shown in Figure 2. In what follows, we give a high-level description of how these text and image features are generated. 


\subsection{Text Embeddings from the Title and Product Description} 

We begin with text embeddings and give a non-technical description of the main ideas.  We will mostly focus on the transformer-based large language model BERT, which is one of the most successful text embedding algorithms.  We give a more technical review of BERT and its predecessors (Word2Vec and ELMO) in Appendix \ref{appendix:text-embeddings}.

\subsubsection{High-Level Objectives}

First, we would like to stress that the high-level objective of the text-embedding algorithms is to construct a concise (low-dimensional) numeric representation of sequences of words, such as product titles and descriptions.

Conceptually, the $j$-th word in the product description can treated as a categorical variable and represented with a very large number $d$ of dummy variables\textemdash where $d$ is the size of the dictionary of words. Still, this representation is not very useful: it is not able to explore word similarity to compactly approximate the dictionary. Indeed each distinct word has a distance of $1$ every other word. 

Instead, we aim to represent words by  vectors of much lower dimension $r \ll d$, such that the distance between similar words is small.
Denote such potential representation of $j$-th word by $u_j$, then the dictionary is $r \times d$ matrix 
$$
\omega = \{ u_{j}\}_{j=1}^d,
$$
where $r$ is the reduced dimensionality of the dictionary, then each word $t_j$ in a  human-readable dictionary can be represented by the word $u_j$.  

Text embedding algorithms aim to find an effective representation of dimension $r$, where $r$ is much smaller than $d$.  This is achieved by treating the entries of $\omega$ as parameters and estimating them
so that the model performs well in certain natural language processing tasks, such as predicting an omitted word in a sentence using surrounding words, or detecting when two sentences are presented in reverse order. Examples are drawn from corpora of published text (ranging from ``small" data, such the entirety of Wikipedia, to a large fraction of all digitized text in the case of GPT-3).  These tasks are not related to hedonic prices\textemdash but precisely because they are not related, one can generate extremely large data sets of examples on which the model can be trained.

Once embeddings for individual words have been obtained, we can generate the embedding for the title or description
of product $i$,  containing the embedded words $\{U_{j,i}\}_{j=1}^J$ by taking averages:
\begin{equation}
W_i = \frac{1}{J} \sum_{j=1}^J \lambda_j  U_{j,i},
\end{equation}
where $\lambda_j$'s are weights given to the $j$-th word. A simple choice is $\lambda_j = 1/J$, but one can also use data-dependent weights (see the Appendix for details).  One may also simply concatenate the embeddings:
$ W_i = \mathrm{vec}( \{ U_{j,i}\}_{j=1}^J).$
This leads to $rJ$-dimensional embedding, but as long as $rJ$
is not overly large, it remains practical.

Once we have obtained the embeddings, how do we judge whether the text embedding is successful? The most obvious check we can do, in our context, is to see if the embeddings are useful for hedonic price prediction. We find that they are extremely useful, as we report later in the paper.  One can also check qualitatively to see if words that have similar meanings have similar embeddings. Ordinarily, this is done through the correlation or cosine-similarity:
$$
\text{sim}(T_k, T_l) =  U_k'U_l/(\|U_k\| \|U_l\|).
$$
According to our human notion of similarity, the more similar the words are (controlling for context), the higher the value the formal similarity measure should take, up to a maximum value of 1. Even the first generation of text embedding algorithms was able to achieve remarkable qualitative performance at this task.
We give such examples in Appendix \ref{appendix:text-embeddings}; the latest generation of the algorithms have only improved in their ability to understand language.  In what follows, we present a non-technical discussion of BERT, a state-of-art language understanding model which we have used extensively; in the Appendix, we present a more technical discussion.

\subsubsection*{BERT at a high level}

BERT (Bidirectional Encoder Representations from Transformers, \citealt{devlin2018bert,vaswani2017attention}) is a transformer-based model developed by Google. The model is trained using a transfer learning approach, meaning auxiliary prediction tasks---predicting a masked word in a sentence or predicting the order of two sentences---that are not connected to the final tasks, such as predicting hedonic prices in our case.  The auxiliary tasks are selected such that the amount of examples is very large (e.g., all of Wikipedia), and such that performing the tasks accurately reflects a high level of comprehension. This initial training with auxiliary tasks is only performed once, so that users may apply the model to smaller datasets without incurring the computational burden of training. Fine-tuning, or further optimization of the network's weights (which is not computationally expensive), is performed on Amazon's product descriptions for apparel, as well as on our final price-prediction task. 

BERT is particularly good at understanding the meaning of words in context, and this property is generally attributed to the transformer blocks present in the neural network \citep{vaswani2017attention}. The transformer blocks in BERT allow the model to understand the context of a word by looking at the words that come before and after it, rather than just relying on the individual word. 
The transformer blocks comprise two main components: a self-attention mechanism and a standard feed-forward neural network.  The self-attention mechanism allows the model to weigh the importance of different words in a sentence when making predictions, producing so-called ``attention weights.'' In fact, BERT uses several self-attention mechanisms in parallel, thus allowing the model to ``attend to'' different parts of the input simultaneously. The feed-forward neural network, also known as a fully connected layer, takes the output from the self-attention mechanism and applies a series of non-linear transformations.  This component allows the model to learn more complex transformations of the input.  Both components are applied to the input in parallel, and then the outputs from both are linearly summarized.  This summary is then used as the input for the next transformer block.


BERT produces text embeddings by breaking the input text up into standardized words or sub-words known as ``tokens.'' These tokens are then passed through the BERT model. The transformer blocks process the tokens in parallel and learn to represent each token in a vector space; this representation is called a word embedding. Indeed, the last hidden layers of the network comprise the embeddings used in our study. During training, BERT learns a set of parameters that can be used to generate embeddings for any input text. The embeddings are then fine-tuned during the fine-tuning phase to perform a specific natural language understanding task; in our case, we fine-tune them on price prediction tasks. The embeddings generated by BERT are contextual, meaning that a word's embedding depends both on that individual word as well as on its context (namely, the other words in the sequence).  This is in contrast to non-contextual embeddings, such as Word2Vec, which assign the same embedding to a word regardless of its context. We present more technical details in Appendix \ref{appendix:text-embeddings}.


It is helpful to compare BERT to GPT (3 and 4), which has received widespread media attention. GPT also uses transformer blocks, but it does so in an autoregressive fashion. It uses supervised learning, which means it is trained on specific tasks to generate human-like text. GPT is pre-trained on a massive amount of text data. It can then be fine-tuned for various natural language generation tasks such as text summarization, text completion, and translation. While GPT is a tool for generating human-like language, BERT is a tool for understanding language. However, since GPT also must understand language in order to generate conversation, future work may investigate whether GPT can provide embeddings that outperform those generated by BERT. 

\subsection{Image Embeddings using ResNet-50}

ResNet-50 is a convolutional neural network architecture designed for image classification \citep{he2016deep}. Convolutional neural networks work by pooling information extracted from sub-images (say, from all $16\times16$ and $4 \times 4$ pixel fragments of a $128 \times 128$ pixel image). 
The key innovation in ResNet is the use of residual connections: shortcut connections that 
act as the identity map, bypassing one or more layers.
Residual connections allow for a very deep network structure (i.e., a highly flexible model) without suffering from vanishing gradients, a problem that typically limits the ability to learn from data. 
Consequently, it becomes possible to train much deeper networks while maintaining good performance. 
ResNet-50 is trained on the ImageNet dataset (with over 14 million images and more than 22,000 different labels). The model has been trained to classify object images into 1000 groups with high accuracy. At the time of its release, the ResNet50 model
achieved the best results in image classification, particularly for the ImageNet and COCO datasets.  There are recent advances in computer vision\textemdash vision transformers\textemdash that import the transformer architecture from large language models such as BERT. They also achieve a near state-of-art performance, but we have not yet explored their use in our context.

Just like with text embeddings, we are not interested in the final predictions of these networks but rather in the last hidden layer, which is taken to be a meaningful summary or ``embedding'' of the image. The reason is that the last hidden layer is used for object classification using a simple logistic model, so it should represent the object type accurately.  This information is clearly useful for our purposes.

\section{AI-Based Hedonic Price Models and Price Indices for Apparel}\label{sec:empirics}

Having constructed numeric embeddings which capture price-relevant characteristics of the product as expressed in its image and text description, we may set about estimating the hedonic price function. In this section, we compare the performance various estimation strategies. We then report derived indices for apparel under our preferred method and contrast it with other major indices.


\subsection{Data}
\label{subsec:data}
We use Amazon's proprietary data on daily average transaction prices and quantities for the first party sales from the entire population of apparel products, with tens of millions of products sold in the Amazon marketplace.\footnote{The quantity information is proprietary, but we note that there are methods of approximating quantity weights based upon product ranking data;
see, e.g., \cite{chessa2019comparing} and \cite{office_of_national_statistics_using_2020}. Therefore,  hedonic prices and hedonic price indices derived using quantity weight can be approximated to various extents by publicly available price information,
product information and images, and product rank information.  }
 Our study covers the period from 01/2013 to 12/2018. The transaction prices of a product $i$  in month $t$ are defined
as the ratio of total sales ($S_{it}$) over the quantity sold ($ Q_{it}$),
$$
P_{it} = S_{it}/Q_{it},
$$
where the price is treated as missing for the case of no sales.

To estimate the hedonic price function, we collected the most recent description (Text$_i$) and image (Image$_i$) available for each product as of July 2019\textemdash shortly after the end of the study. Descriptions include the product's title, brand name, a list of high-level bullet points, and a description provided by the seller.

These images and descriptions are occasionally updated. We remark that changes to images and descriptions do not reflect changes to the underlying product, which would require creation of a new product identifier (see Section \ref{subsec:defining-products} below). Thus, the captured descriptions and images represent each product throughout the study's duration. Changes to the description or image do not reflect changes to a product's underlying valuable attributes, hence they should not change its hedonic price.

\begin{figure}[h!]
\begin{center}
\includegraphics[width=4in]{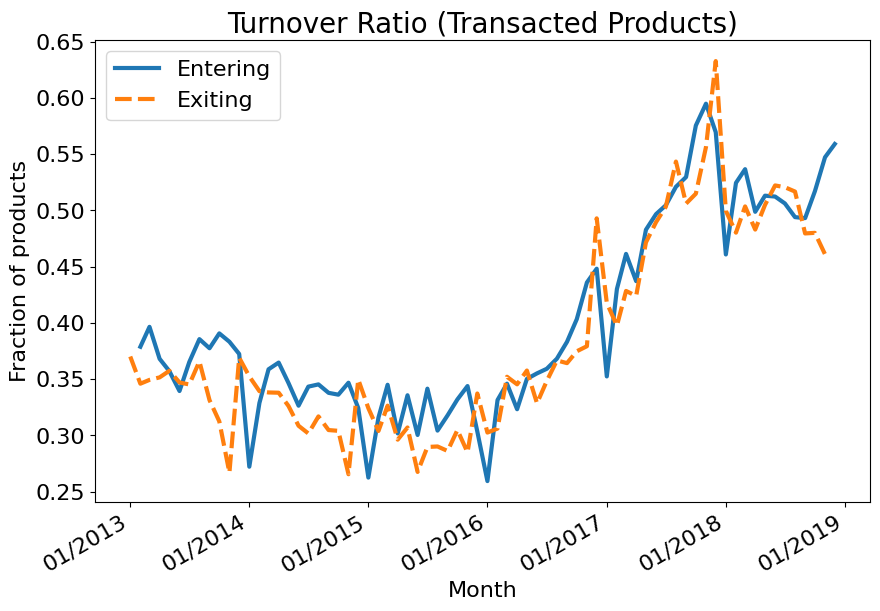}
\par\end{center}
\caption{Turnover Rate for Products. The Figure shows the share
of products with transactions in a given month and no transactions in the previous month (blue line), as well as the share with transactions in a given month and no transactions in the next month (orange line).}\label{fig:turnover}
\end{figure}

\begin{figure}[h!]
\begin{center}
\includegraphics[width=4in]{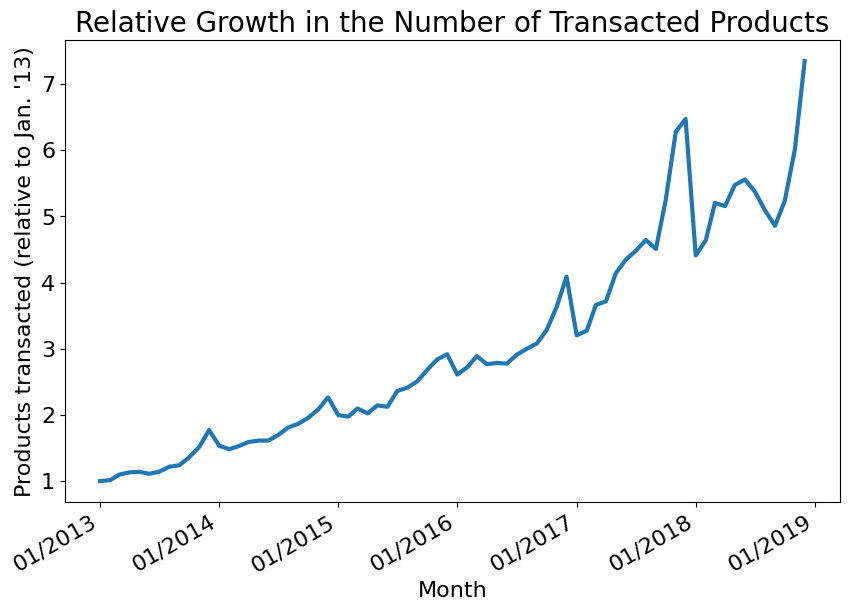}
\par\end{center}
\caption{Products transacted per month relative to the number of products transacted in January, 2013. 
}\label{fig:growthproducts}
\end{figure}

One key characteristic of our data is the very high turnover of products from month to month, as shown in Figure \ref{fig:turnover}.  Moreover, Figure \ref{fig:growthproducts} shows that there is considerable growth in the selection of products. As discussed in the introduction, these properties motivate the use of hedonic indices.  

The size of the resulting dataset is approximately 20 terabytes, owing to the vast number of products and the length of the time period covered. Cloud computing tools based upon Apache's Hadoop and Spark were nonetheless highly effective for holding and processing the data (both in terms of cost and computing time).\footnote{For example, training our model takes under 24 hours on a cloud computer with 8 Nvidia V100 GPUs, corresponding to less then $\$500$ in cloud computing fees using the Sagemaker environment provided by Amazon Web Services. The largest expense was data storage, which came to roughly $\$460$ per month.} In this case, these cloud computing resources were provided by Amazon Web Services (AWS). Computation of the product embeddings and estimation of the hedonic price function were carried out on a GPU cluster, also provided by AWS.

\subsubsection{Defining products} \label{subsec:defining-products} In conducting our study, we have chosen to define ``products'' at a very granular level. In particular, different sizes, colors, and versions of an article of clothing are treated as distinct products, even if they are share the same brand and item name. This gives the neural network model flexibility to group products and adapt to differences in demand between, say, more- or less-popular colors of a given item, and limits the role of human judgement.\footnote{Other choices are possible. See \citet[Sec. 4.1]{aizcorbe2014practical} for further discussion on how researchers may choose to define the product.} This level of granularity is particularly natural in our context, since product variations are typically listed at different prices in Amazon's marketplace. It does, however, influence the interpretation of Figures \ref{fig:turnover} and \ref{fig:growthproducts} as it affects the measurement of product variety and product turnover in our data. Using an auxiliary catalogue of all product variations, we verify in Appendix \ref{subsec:siblings} that the high out-of-sample accuracy reported later in this section is not driven by variations of essentially similar products: we observe similar accuracy when restricting to products from the hold-out sample which are not a variation of any product in the training data.
\subsection{Out-of-sample Performance for Predicting Prices}
\label{subsec:predictive-performance}
In Table \ref{table_R2}, we first examine the predictive performance, recording the $R^2$ for predicting prices in the hold-out sample of products.
In addition to the multi-task price prediction neural network using text and image embeddings $W$ (from BERT) and $I$ (from ResNet50), which is discussed extensively in Sections \ref{sec:model} and \ref{sec:embeddings}, we consider the following alternative methods for estimating the hedonic price function:
\begin{itemize}
\item linear regression using product category and sub-category indicators;
\item linear regression using text and image embeddings $W$ and $I$ generated by BERT and ResNet50 as discussed in Sec. \ref{sec:embeddings};
\item gradient-boosted trees using the aforementioned features $W$ and $I$;\footnote{Gradient-boosted trees (see \citealp{chen2016xgboost}) are a relative of random forests; they have been shown to perform comparably in general prediction tasks and are easier to estimate in very large datasets.}
\item an alternative, ``single-task'' neural network with embeddings $W$ and $I$\footnote{Roughly, single-task learning corresponds to estimating a separate model in each time period.
See Appendix \ref{appendix:alternative-nn-models} for a formal definition.
}
\end{itemize}
Results from the comparison are seen in Table \ref{table_R2}. When we switch from the basic catalog features to the embeddings $W$ and $I$, we obtain a first major improvement\textemdash even using linear regression in the final step. We obtain a second major improvement when we switch from linear regression to a scalable implementation of tree-based methods (gradient-boosted trees). Finally, we obtain the final major improvement as we switch from tree-based methods to a multi-task neural network. The neural network model achieves substantially better predictive performance than other methods.
\begin{table}[t!]
 \begin{center}
\begin{tabular}{lcc}
\hline\hline
Method  & &R$^2$  \\
\hline
Linear Model with basic catalogue features   &   $\approx$  & $30-45\%$ \\ 
\hline 
Linear Model with embeddings $W$ and $I$ &   $\approx$   & $55-65\%$ \\ 
Random Forest/Boosted Tree Models with embeddings $W$ and $I$ &  $\approx$ &  $70-80\%$  \\
\hline
Single-Task Neural Network with embeddings $W$ and $I$  &  $\approx$ &  $75-85\%$ \\
Multi-Task Neural Network with embeddings $W$ and $I$ &  $\approx$ &  $80-90\%$\\
\hline\hline
\end{tabular}
\caption{Summary of Out-of-Sample Performance of the Empirical Hedonic Price Function.}
\label{table_R2}
\end{center}

\end{table}

\begin{figure}
\begin{center}
\includegraphics[width=4in]{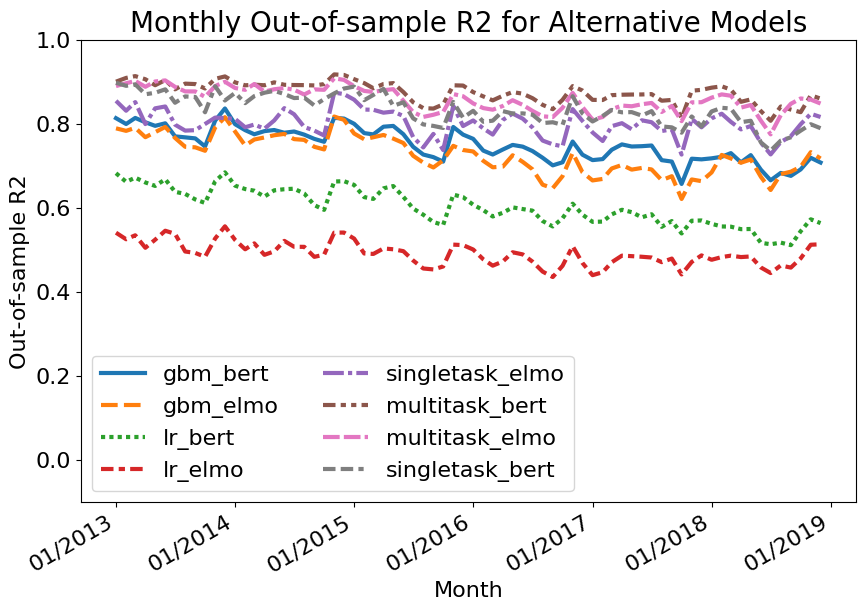}
\end{center}
\caption{The out-of-sample performance of the empirical hedonic price function obtained using our neural network every month since January, 2013, in comparison to alternative models. Multi-task neural networks
dominate single-task neural networks (discussed in Appendix \ref{appendix:alternative-nn-models}). Neural network models dominate boosted tree models (GBM),
which in turn dominate linear models (LM).}
\label{fig_r2apparel}
\end{figure}

Figure \ref{fig_r2apparel} also presents the month-to-month performance of the various models. The out-of-sample  $R^2$
for the best multi-task neural network model ranges between $80\%$ to $90\%$. Multi-task neural networks (NNs)
uniformly dominate single-task NNs, which in turn uniformly dominate boosted tree models and linear models. We also contrast the performance of BERT-based embeddings of the product description, $W$, with an alternative embedding of the product description produced using the ELMO network\footnote{ELMO is a large language model based upon a recursive neural network architecture. Although it does not use the modern transformer architecture, it performs nearly as well as BERT in our setting. We review details of ELMO in Appendix \ref{appendix:text-embeddings}}
The BERT-based multi-task NN almost always outperforms the ELMO-based
multi-task NN, although the difference in performance is relatively small.

A limitation the multi-task model in comparison to single-task models, however, is that it requires retraining the full model as new data becomes available over time. Moreover, the performance of the best multi-task models is best at the beginning of the studied period and worsens over time, possibly due to the increasing variety and number of products being transacted (shown in Figure \ref{fig:growthproducts}). 
It is worth mentioning that the out-of-sample $R^2$ agrees with validation $R^2$ we obtained in training (not reported), which suggests that our training approach successfully limits overfitting.

\subsubsection{Examples of hits and misses} 
\begin{figure}[h]
\begin{center}
\includegraphics[height=2in]{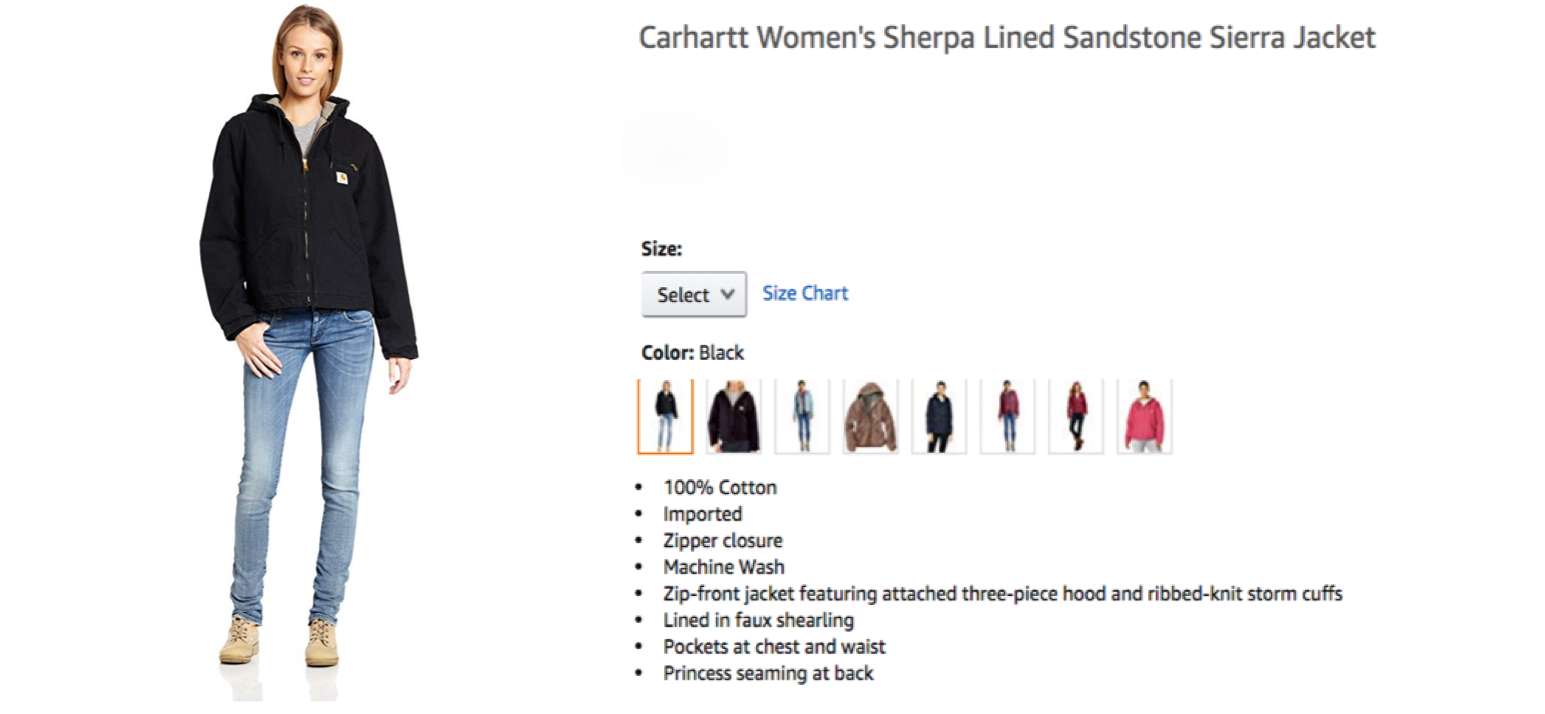}
\par\end{center}
\caption{An example of accurate price prediction for a women's jacket (black, size medium): The neural network  predicts the price of this item at about $\$100$. The average price on the price aggregator \texttt{camelcamel.com} is $\$97$, with the offer price ranging from $\$39$ to $\$120$.}
\label{fig:hitexample}
\end{figure}

\begin{figure}[h]
\begin{center}
\includegraphics[height=2in]{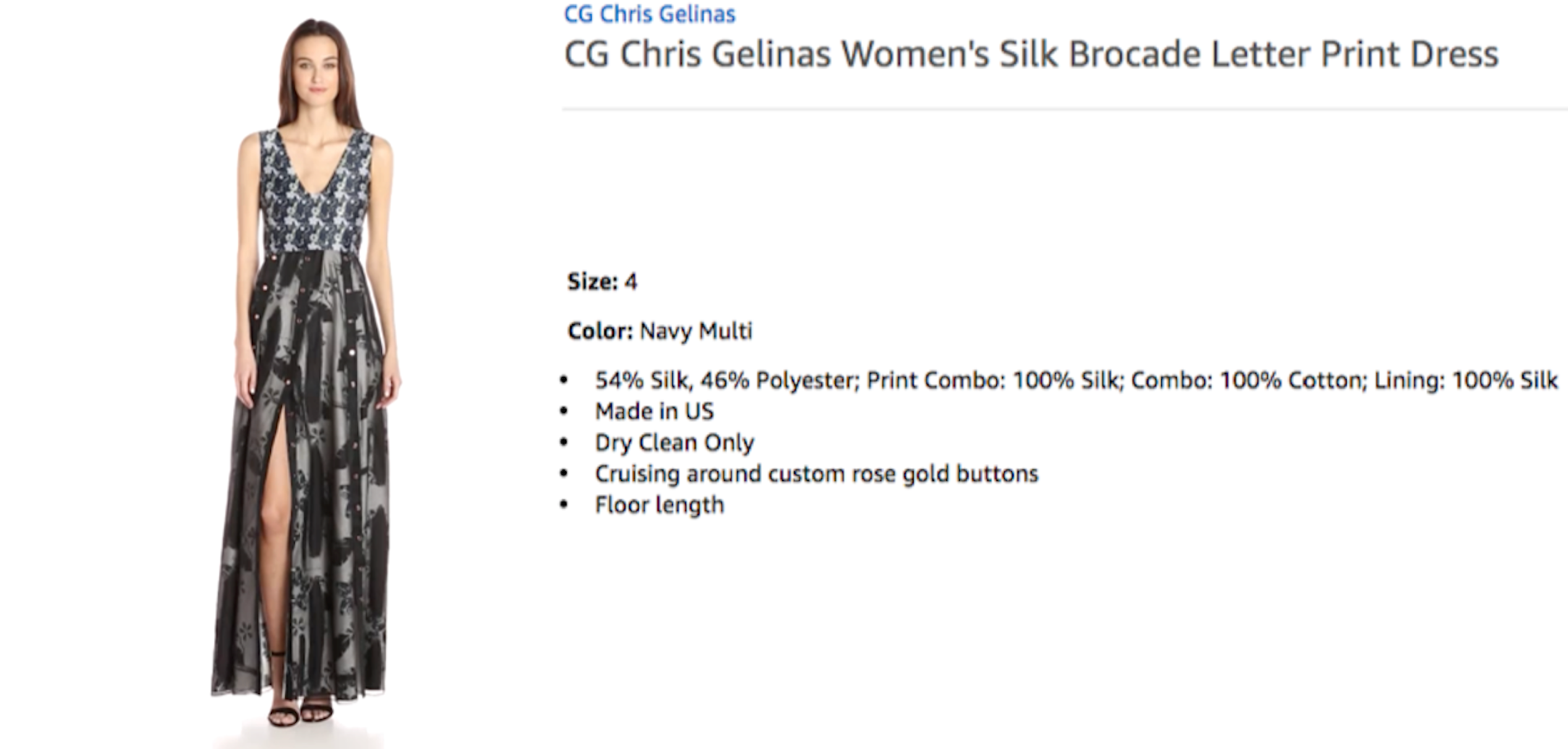}
\par\end{center}
\caption{An example of inaccurate prediction: The neural network predicts the price of this item at about $\$300$, but the true price is roughly $\$2400$.  However, the price history for this item 
as recorded by the price aggregator \texttt{camelcamel.com}, shows that its price ranged between $\$206$ and
$\$2800$, with an average price of $\$464$. }
\label{fig:missexample}
\end{figure}

We can inspect the performance of the best predictive model using examples. In Figures \ref{fig:hitexample} and \ref{fig:missexample}, we present one example of accurate prediction and one of inaccurate prediction. For the first item, a sweater, the neural network model predicts a price of about $100$. The time-averaged average price as shown on \texttt{camelcamel.com}\textemdash a third-party website which tracks prices on Amazon\textemdash is $\$97$, with the price ranging from $\$39$ (corresponding to liquidation events) to $\$120$. For the second example, a designer dress, the neural network predicts the price of this item at about $\$300$, but the most recent offer prices for this item were around $\$2400$. While this seems to be a serious inaccuracy, the price history for this item, again per \texttt{camelcamel.com}, suggests that there were periods when the price ranged between $\$206$ and $\$2800$; the average listed price was $\$464$, which is not as far off from the prediction. 

\subsection{Statistical Significance and Inference}
\label{subsec:testing}
\begin{figure}[h]
\begin{center}
\includegraphics[height=3in]{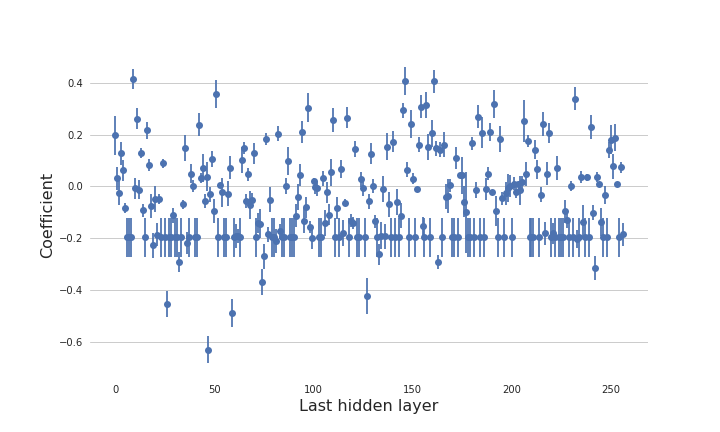}
\end{center}
\caption{
Point estimates and pointwise $95\%$ confidence intervals for linear coefficients of the value
embeddings, as estimated by a linear regression model in the hold-out sample.
Of the 256 coefficients (corresponding to entries of the 256-dimensional value embedding $V_i$), the majority are significant at the $10^{-5}$ level.
}
\label{fig:significance}
\end{figure}

We examine the hedonic price model's statistical significance using the methods in Section 3.2.\footnote{Note that the procedure  involves re-estimating the final model weights, $\theta$. Thus, this section considers slightly different coefficient and hedonic price estimates than those discussed elsewhere in the paper.}  Figure \ref{fig:significance} reports the estimated coefficients on value embeddings for the month of November 2018 as well as the component-wise confidence intervals. We also illustrate the construction of the confidence intervals for predicted hedonic price in Table \ref{table:CI}, which reports the $90\%$ confidence intervals for estimated hedonic prices of two example products.

{
\medskip

\begin{table}[h]

\scriptsize
\begin{tabular}{p{3cm}|ccccccccc}
\hline\hline\\
 Product & Month	& $P_{it}$	 & $\widehat H_{it}$	& $  \mathrm{SE}_{it}$ & $[L_{it}, U_{it}]$   & $[L_{it}(P_{it}),\,U_{it}(P_{it})]$	 \\
\hline
\\
 W M Black Jacket 
 & Dec. 2018	& 118.8	 & 114.6   &0.05	  & [114.5, 114.7]	 & [102, 126]\\
 W XS Blue Jacket 
 & Dec. 2018	& 119.9	& 110.25	 & 0.06 &  [110.1,	 110.4] & [98 , 122]\\
\\
\hline\hline
\end{tabular}
\caption{\normalsize Confidence Intervals for Predicted Hedonic Prices $H_{it} = V_{it}'\theta_t$
and Sale Prices $P_{it}$.  Here,
$\widehat H_{it} = V_{it}'\hat \theta_t$ is the estimated hedonic price. The term $\hat \sigma^2=V_{it}' \widehat{\mathrm{Cov}}(\hat \theta_t) V_{it} $ is the square of the standard error, and $[L_{it},U_{it}] = [\widehat H_{it}  \pm z_{.95}\hat \sigma]$ is the $90\%$ confidence interval for $H_{it}$, where $z_{.95}$ is the $95^{\text{th}}$ percentile of the standard normal distribution..  The predictive confidence interval for $P_{it}$ is $[L_{it}(P_{it}),U_{it}(P_{it})]=[\widehat H_{it} \pm z_{1-\alpha/2}  \hat  \nu]$ with $\hat \nu^2 = \hat \sigma^2 + \widehat \Var ( P_{it} - H_{it} )$.}
\label{table:CI}
\end{table}}

\subsection{Hedonic Price Indices for Apparel}

We will now use our price predictions to construct hedonic indices, following the definitions introduced in Section 2. For this exercise, we use the best predictive model from the preceding comparison, which we have described in detail in Sections \ref{sec:model} and \ref{sec:embeddings}.

In Table \ref{table:inflation}, we present our main results\textemdash estimates of the average annual rate of inflation in apparel from 2014 to 2019, via Fisher hedooic price indices (yearly-chained, monthly chained, and the geometric mean of the two), the Jevons posted-price index, the Adobe DPI \citep{adobe2024digital}, and CPI \citep{bls2024cpi}:
\begin{table}[h!]
\begin{center}
\begin{tabular}{lc}
\hline 
\hline
\textit{Apparel Indices} & \textit{Change in Price Index}\tabularnewline
\hline 
\textbf{Fisher Hedonic, Yearly Chaining (FY)} &  -0.98\% \tabularnewline
Fisher Hedonic, Monthly Chaining (FM) &  -5.27\% \tabularnewline
Fisher Hedonic, Geometric Mean ($\sqrt{FY \cdot FM}$) & -2.28\% \\
Fisher Matched, Monthly Chaining (FI) & -3.12\% \tabularnewline
Jevons Posted Price Index, Daily Chained (JPI) & -3.01\%\\
Adobe Digital Price Index, Monthly Chained  (DPI) & -2.02\% \\
\textbf{U.S. Urban Apparel (BLS)} &  -0.31\% \tabularnewline
\hline 
\hline
\end{tabular}
\end{center}

\caption{Estimates of Average Annual Rate of Inflation in Apparel over five years, 2014-2019:
Fisher Hedonic Index, Fisher Matched Index, Jevons Posted Price Index, Adobe DPI, and 
the BLS urban CPI.}\label{table:inflation}
\end{table}

The main conclusions we draw from these results are the following.

\begin{itemize}

\item[E.1]  The main index -- the yearly-chained FHPI -- suggests an average price decline in 2014-2019 of $.98\%$. In contrast, the BLS CPI for apparel suggests that there was a somewhat smaller decline in the price level. \\

\item[E.2] The yearly chained FHPI declines at a slower rate than the matched Fisher index, Adobe DPI, and the monthly-chained FHPI, highlighting the importance of using hedonic adjustment and long chaining. \\

\end{itemize}
Indeed, the matched index potentially contains selection bias induced by the rapid turnover of products, illustrated in Figure \ref{fig:turnover}.  Moreover, as we emphasized in the introduction, the use of hedonic prices allows us to perform ``long" chaining, thereby mitigating the chain drift problem.

Differences from the CPI could be attributed to a number of sources:  There are methodological differences: the BLS uses a hybrid index where price levels are first measured within narrow subgroups of products, without quantity weighting, and then aggregated using a Tornqvist index with  expenditure shares for subgroups. Moreover, the BLS also uses their own hedonic models to perform quality adjustments.\footnote{Note that our goal here is also not to replicate the BLS CPI index but to construct a modern hedonic version of a classical Fisher index. We also note that the replication is simply precluded in our setting because we can not assign extremely many products manually to the subcategories used by BLS.} \cite{moulton2018measurement} estimates that the upward bias in the overall (non-apparel specific) CPI index may be in the range $[.4\%, 1.3\%]$,  which could potentially reconcile some of the difference.  Other sources of the difference could be the Amazon-specific productivity improvements leading to lower prices (e.g., improvements in supply-chain productivity) and different shares of products in customers’ baskets.

We also performed similar calculations using a linear model instead of the neural network (not reported), and 
conclusions are qualitatively very similar. However,
the neural network models\textemdash which do exhibit superior predictive ability\textemdash result in a more pronounced quantitative drop
in the price index level. The differences between the two are relatively small, likely due to the fact that quantity-weighted averaging eliminates noise present in the linear model's predictions. 

\begin{figure}
\begin{center}
\includegraphics[width=4in]{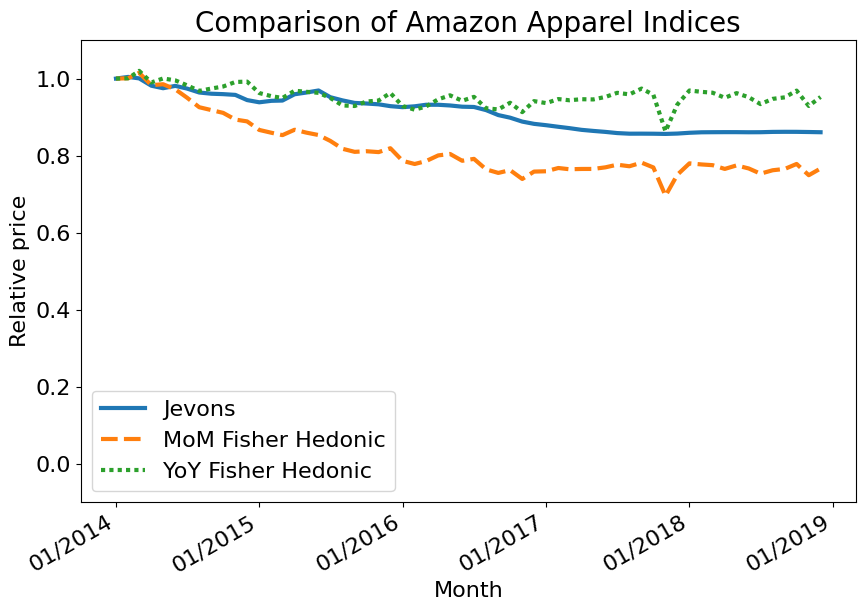}
\par\end{center}
\caption{Dynamics of the yearly-chained Fisher Hedonic Price Index (dotted line),  the Jevons Price Index (solid lined), and  the monthly-chained Fisher Hedonic Price Index (dashed line).}
\label{GEKS}
\end{figure}

In Figure \ref{GEKS}, we demonstrate the dynamics of the year-chained FHPI and compare it with the Jevons index and the monthly chained FHPI. 

The Jevons index is a geometric mean of the posted price relatives and does not incorporate quantity weighting. It underlies the Billion Price Project \citep{cavallo2018scraped} and is also used by statistical agencies as a complementary index \citep{office_of_national_statistics_using_2020}.  The Jevons index is convenient because it only relies on publicly posted prices and does not use quantity information; it measures within-product price level changes which are averaged over a much larger universe of products (with and without transactions). In contrast, the Fisher index reflects both within-product price changes and the as well as changes due to substitution arising from utility-maximizing behavior by customers with budget constraints.  The key empirical observations are as follows.

\begin{itemize}

\item[E.3] The Jevons index exhibits steady declines over the studied period, with a total decline of more than $10\%$; the rate of decline is considerably larger and much more implausible than that of the yearly chained FHPI.

\end{itemize}

This may suggest that either the chain drift or the substitution effects are large enough to matter qualitatively. Without the ability to do long chaining and 
without any proxy for quantity information (such as sales rank data), it is doubtful that Jevons-type indices can accurately approximate inflation. Still, the Jevons index continues to be a very important, simple way to gauge large changes in inflation, as the BPP (based on the Jevons index) has amply demonstrated (\cite{cavallo2016billion}). \footnote{Note that given a reference series, such as the CPI, one can always use a post-processing or ``nowcasting" approach to modify Jevons or any other digital index $R_g$ to track the reference index $R_f$ as well as possible. For example, one can construct the best linear predictor $\hat a + \hat b \log R_g$ of $\log R_f$ using the historical data on $(R_g, R_f)$ in a given time frame and then use $ R^\star_g = \exp(\hat a + \hat b \log R_g)$ as the post-processed index. This further highlights the potential usefulness of the Jevons and other digital indices, even though their raw versions exhibit large biases. In our discussions, we focus on raw indices without post-processing, with the understanding that post-processing can always be employed as needed. }

Finally, when we compare yearly-chained FHPI to monthly-chained FHPI, the key observation is as follows:

\begin{itemize}
\item[E.4]  The monthly-chained  FHPI exhibits a strong, very implausible $20\%$ decline in the price level, in sharp contrast to the yearly-chained FHPI. 
\end{itemize}

This  highlights the importance of mitigating ``chain drift,"  which appears to be quite severe for frequently compounded series such as the monthly chained FHPI.  For this reason, the monthly chained FHPI and other frequently chained indices (matched Fisher and daily-chained Jevons) are unlikely to measure inflation accurately, especially over longer horizons.  Of course, as we stressed before for the Jevons index, these indices remain useful over short horizons to gauge large changes in the price level.\footnote{They can also be employed in a nowcasting fashion, as explained in the previous footnote.}

\section{Conclusion}

 We develop empirical, AI-based models of hedonic prices, and use them to derive hedonic price indices for measuring changes
in consumer welfare. To achieve this, we generate product attributes (or 'features') from text descriptions and images using deep neural networks, namely BERT (for text) and ResNet-50 (for images). We then use these features to estimate the hedonic price function---again using deep learning---via a multi-task neural network that predicts a complete time-series of prices for each product.  
All the ingredients to the method rely on publicly available, open-source software components. 
We apply the models to Amazon's proprietary data on first-party sales in apparel. Resulting models have high predictive accuracy for several product categories, with the $R^2$ ranging from $80\%$ to $90\%$. Our main hedonic index estimates the rate of inflation in apparel to be moderately negative, somewhat lower than the rate estimated by the CPI.

We find the performance of AI-based hedonic models remarkable for two reasons. First, their high predictive accuracy suggests that theoretical hedonic price models provide a good approximation of real-world equilibrium prices. Second, our methods are scalable to many products and avoid the significant manual effort required to construct more traditional hedonic price indices. The good performance of these methods suggests that AI-powered methods, particularly our embeddings, can be used to build real-time hedonic price indices using electronic data and facilitate price research in many settings. These methods can also power economic research in many other areas\textemdash for example, labor economics, where embeddings can be used to characterize workers' job-relevant characteristics, or industrial organization, where embeddings can be used to characterize firms and their products. Since all main ingredients of our approach are open-source, publicly-available technology, future scientific pursuits in this direction can readily use these tools or other similar open-source products.

We believe there are natural directions for further research. First, we found that while images used alone help predict prices, they add very little to prediction accuracy once the text embeddings are included in the model. Therefore, finding better ways to leverage image data is an important unresolved problem. Second, an important further research direction is model explainability. When the price of the product changes, we'd like to explain the price change in terms of changing valuations of key attributes. Similarly, when comparing the prices of two products, we would like to attribute the price difference to the valuations of key attributes. Given the non-linearity of the AI-based hedonic models, explainability is not a simple problem. We hope interested researchers take notice of these open problems.
\section{Acknowledgements}

We thank the participants at the
2018 Federal Economic Statistics Advisory Committee meeting, the 2019 Allied Social Science meetings, the 2019 Federal Reserve Board conference on "Nontraditional Data, Machine Learning, and Natural Language Processing in Macroeconomics", the 2019 Brookings Conference "Can big data improve economic measurement?" and the 2021 CEMMAP conference "Measuring prices and welfare,"  and seminars at Berkeley, European Bank of Reconstruction and Development, MIT, UCL, and York.  We are grateful to Andrew Chesher, Greg Duncan, Kevin Fox, John Haltinwager, James Heckman, Daniel Miller, Mathew Shapiro, Bernhard Schölkopf, James Stock, and Weining Wang  for helpful comments during the various stages of this project. All authors were fully or partially employed by Amazon.com while conducting the research. 

\bibliographystyle{plainnat}
\bibliography{bibliography}

\newpage

\appendix

\clearpage

\section{Overview of Text Embedding Models}
\label{appendix:text-embeddings}
Here we provide a more technical overview of the text embedding models used in this paper.

\subsection{First generation: Word2Vec Embeddings} We first recall some basic ideas underlying the Word2Vec algorithm \citep{mikolov2013efficient}. Here we rely on notation in Section 4. The goal is to find the $r \times d$ matrix 
$
\omega = \{ u_{j}\}_{j=1}^d,
$
representing $d$ words in $r$ dimensional space. The columns are the embeddings for the words.


We can think of a word appearing in sentence as random variable $T$; and we can let $U$ denote the corresponding embedding. Word2Vec trains the word embeddings by predicting the middle word from the words that surround it in word sentences.  Given a subsentence $s$ of $K+1$ words, we have a central word $T_{c,s}$ whose identity we have to predict and we have the words $\{T_{o,s}\}$  that surround it.  Collapse the embeddings for context words by a sum,
$$
\bar U_o = \frac{1}{K}\sum_{o} U_{o,s},
$$
where $U_{o,s}$ is the element of $\omega$ corresponding to the word $T_{o,s}$. This step imposes a drastic simplifying assumption that the context words are exchangeable.

The probability  of middle word $T_{c,s}$ being equal to $t$ is modeled via multinomial
logit function:
$$ 
p_{s} (t; \pi, \omega):= P\Big (T_{c,s} =t\mid\{T_{o,s}\}; \omega \Big ) = \frac{\exp(\pi_t' \ \bar U_s(\omega))}{\sum_{\bar t} \exp({\pi_{\bar t}'  \ \bar U_s(\omega)}) },
$$
where $\pi = (\pi_1,... ,\pi_d)$ is $m \times d$ matrix conformable parameter vectors defining the choice probabilities.  
The model constraints $\pi = \omega,$ and estimates $\omega$ by using the maximum quasi-likelihood method:
$$
\max_{\omega = \pi} \sum_{s \in \mathcal{S}}  \log p_s(T_{i,s}; \pi, \omega),
$$
where the summation is over many examples $\mathcal{S}$ of subsequences $s$. r

In summary, the Word2Vec algorithm transforms text into a vector of numbers that can be used to compactly represent words. The algorithm trains a neural network in a supervised manner such that the contextual information is used to predict another part of the text. For example, let's say that the title description of the item is: ``Hiigoo Fashion Women's Multi-pocket Cotton Canvas Handbags Shoulder Bags Totes Purses". The model will be trained using many $n$-word subsentence examples, such that the center word is predicted from the rest. If we just use $K=3$ subsentence examples, then we train the model using the following examples: (Hiigoo,Women's) $\rightarrow$ Fashion, (Fashion,Multi-pocket)$\rightarrow$ Women's, (Women's,Cotton) $\rightarrow$  Multi-pocket, and so on.  

We can examine the quality of word embedding by assessing predictive performance for price prediction tasks.  We can also qualitatively inspect whether embeddings capture word similarity. For example, we found that the embeddings for ``necktie" and ``bowtie" are most cosine-similar to the word ``tie." The embeddings also seem to induce an interesting vector space on the set of words, which seems to encode analogies well. For example, the embedding for the word ``briefcase" is most cosine-similar to the artificial latent word $$ \mathrm{Word2Vec(men's)} +  \mathrm{Word2Vec(handbag) }-  \mathrm{Word2Vec(women's )}. $$
Examples like this  and others reported in \cite{mikolov2013efficient} supported the idea of summing the embeddings for words in a sentence to produce an embedding for sentences. 

Word2vec embeddings were among the first generation of early successful algorithms. These algorithms
have been improved by the next generation of NLP algorithms, such as ELMO and BERT,
which are discussed next.

\subsection*{Second Generation: ELMO} 
The Embeddings from Language Models (ELMO) algorithm (Peters et al, 2018) uses the ideas of the Shannon game, where we guess the next word in the sentence $m$ with $n$ words, i.e. 
$$p_{k,m}^f (t) =  P[T_{k+1,m}=t | T_{1,m}, ..., T_{k,m}; \theta]$$
and also uses reverse guessing as well:
$$p_{k,m}^b (t) = P[T_{k-1,m} =t | T_{k,m}, ...T_{n,m}; \theta],$$
where $\theta$ is a parameter vector. Recursive neural networks with single or multiple hidden layers are used to model these probabilities.  Parameters
are estimated using quasi-maximum log-likelihood methods, where the forward and backward log quasi-likelihoods are added together.

To give a simple example, suppose we wanted incorporate more contextual information into our word embeddings. Instead of collapsing embeddings for the context word by a sum, we could assign individual parameters to each preceding word. This would 
result in a model closely resembling the previous model, but where the order of context words would play a role. For example, we could model
$$ 
P(T_{k,m} = t \mid\{T_{j,m}\}_{j=1}^{k-1}) = \frac{e^{\sum_{j=1}^{k-1} \pi_{t,k}'U_{k,m}(\omega)}} {\sum_{\bar t} e^{ \sum_{j=1}^{k-1} \pi_{\bar t,k}'U_{k,m}(\omega)}}  ,
$$
and similarly in reverse order. ELMO uses a more sophisticated (and more parsimonious) non-linear recursive  nonlinear regression model to build these probabilities, shown in Figure \ref{elmo}. The resulting model is an example of a recurrent neural network (or RNN).


The basic structure of ELMO is as follows: Given a sentence $m$ of $n$ words, (1) words are mapped to context-free embeddings in $\mathbb R^d$. (2) A network is trained to predict each word $T_{k,m}$ of a string given (a) words $(T_{1,m}, \ldots,T_{k-1,m})$ or (b) words $(T_{k+1,m}, \ldots, T_{n,m})$. The objective is to minimize
the average over the sum of the log-loss over the $2n-2$ prediction tasks, where the average is taken over all sentences.
(3) The embedding of word $T_{k,m}$ is given by a weighted average of outputs of certain hidden neurons corresponding to this word's entire context. Importantly, the same final  logistic (``softmax") layer is used for prediction objectives (2a) and (2b). Thus the inputs to this layer, which represent the forward and backward context, are constrained to lie in ``the same space.''

\begin{figure}
\begin{center}
\subfile{figures/elmo-diagram}
\end{center}
\caption{ELMO Architecture. This is an ELMO network for a string of $4$ words, with $L=2$ hidden layers. Here, the softmax layer (multinomial logit) is a single function mapping each input in $\mathbb R^d$ to a probability distribution over the dictionary $\Sigma$.}\label{elmo}
\end{figure}
\subsubsection{Training} In Figure \ref{elmo}, the output probability distribution $p^{f}_k$ is taken as a prediction of $T_{k+1,m}$; similarly $p^b_k$ is taken as a prediction of $T_{k-1,m}$. The parameters of the network $\theta$ are obtained by 
maximizing the quasi-likelihood:
$$\max_\theta \sum_{m \in \mathcal{M} } \left( \sum_{k=1}^{n-1} \log p_{k,m}^f (T_{k+1,m}; \theta) + \sum_{k=2}^{n}  \log p^b_{k,m}(T_{k-1,m};  \theta) \right), $$
where $\mathcal{M}$ is a collection of sentences (titles and product descriptions)  in our data set.

\subsubsection{Producing embeddings}
To produce embeddings from the trained network, each word $t_k$ in a sentence $m=(t_1,...,t_k)$  is mapped to a weighted average of the outputs of the hidden neurons indexed by $k$:
\[t_k \mapsto w_k := \sum_{i=1}^L \gamma_i w^f_{ki} + \bar \gamma_i w^b_{ki}.\] 
The embedding for the sentence (or product description)  is produced by summing the embeddings for each individual word.
The weights $\gamma$ and $\bar \gamma$ can be tuned by the neural network performing the final task. In principle, however, the whole network can be merged with the network performing the final task and jointly optimized.

\subsection{Second generation: BERT}
Bidirectional Encoder Representations from Transformers (BERT) is another contextualized word embedding learned from deep language model (Devlin et al, 2018).  It is a successor of ELMO and achieved state-of-art results on multiple NLP tasks, improving somewhat on ELMO. Instead of using Recurrent Neural Network as in ELMO, BERT uses the Transformer structure with attention mechanism (Vaswani et al., 2017) that pays attention to whole sentence or context.

Unlike the language model in ELMO which predicts the next word from previous words, the BERT model is trained on two self-supervised tasks simultaneously:
\begin{itemize}
\item Mask Language Model: randomly mask certain percentage of the words in the sequence and predict the masked words
\item Next Sentence Prediction: given a pair of sentences, predict whether one sentence proceeds another.
\end{itemize}

The structure of the BERT model is as follows: (1) Each word in the input sentence is broken to standardized word fragments called ``tokens.''  (2) For each token, its input representation consists of i) a binary encoding of the token from (1), and  ii) a  position embedding indicating the position of the token in the sentence (3) The input representation of tokens in the sequence is fed into the main model architecture: $L$ layers of Transformer-Encoder blocks. Each block consists of a multi-head attention layer, followed by a feed forward layer. 

To initially train the network, a special token $\mathtt{[cls]}$ is added to the beginning of the sequence, and a fraction of the tokens are replaced by $\mathtt{[mask]} at random$. The output representation of the mask token $\mathtt{[mask]}$ is used to predict the masked word via a softmax layer, and the output representation of the special $\mathtt{[cls]}$ token is used for next sentence prediction. The loss function is a combination of the two losses. Once trained, the final hidden layer (excluding the softmax layer used in prediction) is used as the sequence of word embeddings.

We next focus in detail on the main structure in step (3), especially the ``multi-head attention" layer.
\subsubsection{Computing The Attention} We begin with $n$ word embeddings $(x_1, x_2, \ldots, x_n)$, with each $x_k \in \mathbb R^d$. Let $\mathsf{X}$ denote the matrix whose $k$th row is $x_k$.  The Multi-Head Attention mapping is applied on $\mathsf{X}$ directly: $$\mathsf{X} \longmapsto \mathrm{MultiHead}(\mathsf{X} , \mathsf{X} , \mathsf{X} ),$$
where 
$$\mathrm{MultiHead}(\mathsf{Q}, \mathsf{K}, \mathsf{V}) = \mathrm{Concatenate}(\mathrm{Head}_1, \dots, \mathrm{Head}_h)\omega^O,$$
$$\mathrm{Head}_i = \mathrm{Attention}(\mathsf{Q} \omega_i^Q, \mathsf{K}\omega_i^K, \mathsf{V}\omega_i^V),$$
$$ \mathrm{Attention}(\mathsf{\tilde Q},\mathsf{ \tilde  K}, \mathsf{ \tilde V}) = \mathrm{softmax}\left (\mathsf{ \tilde  Q} \mathsf{ \tilde  K}^T/\sqrt{d_k}\right) \mathsf{ \tilde V},$$
where $\omega^O$ and $(\omega_i^Q,\omega_i^K, \omega_i^V)$ are matrix parameters, which are optimized to maximize
the model performance. In other words, each word embedding is replaced by a weighted average of embeddings for all other words, and the weights are learned from the scaled dot-product of different projections of the word embeddings themself. The projection matrices are parameters learned during training. 

\begin{figure}
\subfile{figures/bert-diagram}
\caption{\label{fig:bert}BERT Architecture}
\end{figure}

\subsubsection{Generating product embeddings}
Depending on specific tasks and resources,  Devlin et al. (2018) suggested using the BERT embeddings in various ways: 1) use the last layer, second-to-last layer or concatenate last 4 layers of the encoder outputs from the pre-trained BERT model, 2) fine tune the whole BERT model on the downstream task, or 3) train the BERT language model from scratch on the new data. 
For this study, we choose the feature-based approach and extract the second-to-last layer as embeddings from the pre-trained BERT model. Each product's text embedding is the average of the embeddings of each word/token from the input text field. 

\subsection{Comparing ELMO and BERT}
While ELMO and BERT both represent recent breakthroughs in natural language processing, the former was an initial, highly effective, contextual word embedding model trained using deep learning, while the latter was the first contextual word embedding to fully exploit the Transformer architecture. Since the BERT paper was published second and could respond directly to the ELMO paper (but not vice-versa), its comparisons may be somewhat biased towards BERT.

There are several technical differences between the two proposals:
(1) They use different initial context-free embeddings. ELMO applies an initial convolutional layer to a \emph{character} embedding, while BERT augments a binary encoding of sub-words (i.e.,~tokens) with positional data.
(2) ELMO is based on the recurrent neural network architecture depicted in Figure \ref{elmo} while BERT is based on Transformer architecture in Figure \ref{fig:bert}. 
(3) The ELMO implementation only allows the averaging weights to be fine-tuned, whereas BERT proposes fine-tuning the whole network. Of these, the biggest difference lies in the choice of the model architecture. The transformer architecture appears to be more effective than the RNN at capturing long-range dependencies in text. Furthermore, ELMO creates context by using the left-to-right and right-to-left language model representations, while in BERT 
models the entire context simultaneously.


\section{Overview of Image Embeddings via ResNet50}

\begin{figure}
\begin{center}
\includegraphics[width=6in,height=3in]{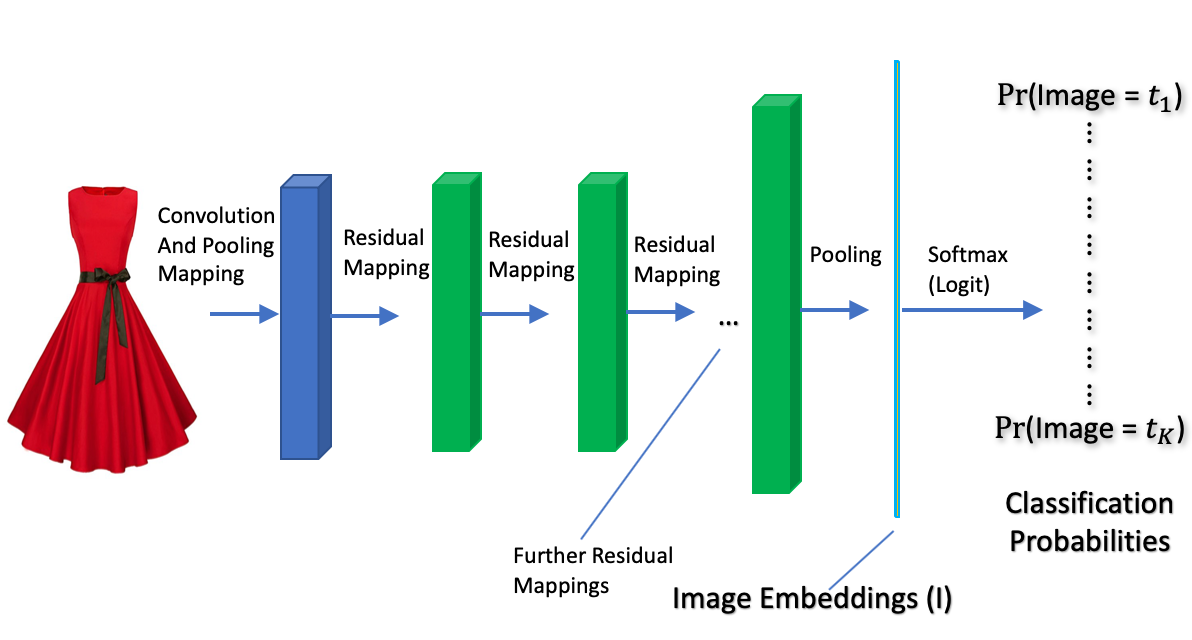}
\par\end{center}
\caption{\small The ResNet50 operates on numerical 3-dimensional arrays representing
images. It first does some early processing by applying convolutional and pooling filters, then it applies many residual block mappings,  producing arrays shown in green. The penultimate layer produces a high-dimensional vector $I$, the image embedding, which is then used to predict the image type. }\label{resnet}
\end{figure}

The central idea of the ResNet is to exploit ``partial linearity": traditional nonlinearly generated neurons are combined
(or added together) with the previous layer of neurons.  More specifically,  a building block 
is to take a standard feed-forward convolutional neural network and add skip connections that bypass two (or one or several) convolution layers at a time. Each skipping step generates a residual block in which the convolution layers predict a 
residual.  Formally each $k$-th residual block is a neural network mapping
$$
 v\longmapsto (v, \sigma^0_k(\omega^{0}_kv)) \longmapsto (v,  \sigma^1_k \circ \omega ^1_k \sigma^0_k(\omega_k v)) \longmapsto v+ \sigma^1_k \circ \omega^1_k \sigma^0_k(\omega^0_k v),
$$ 
where $\omega$'s are matrix-valued parameters or ``weights". This can be seen as a special case of general design pattern.
Putting together many blocks like these sequentially,  we obtain the overall architecture depicted in Figure  \ref{resnet}.

Convolutional Neural Networks (CNNs) are a type of deep learning model especially effective for analyzing visual imagery. They work by using convolutional layers that apply filters to input data, capturing spatial hierarchies and patterns like edges, textures, and shapes. CNNs efficiently learn features from images, making them widely used in tasks such as image classification, object detection, and facial recognition.

Initial prototypes of deep CNNs (those with many layers of neurons) were challenging to optimize: at a certain depth,  additional layers often resulted in much higher validation and training errors. The residual network architecture addressed this by using the residual block architecture outlined above. This allowed for
effective optimization even for very deep networks. 

Just like with text embeddings, we are not interested in the final predictions of these networks but rather in the last hidden layer, which is taken to be the image embedding.  

\section{Details of Different Architectures for Prediction Using Embeddings} \label{appendix:alternative-nn-models}

Here we summarize different configurations of neural networks that we considered.

\subsection{Model 1: ELMO + Single Task Models} 

The first neural network model we tried is the Single Task model, i.e. predicting product prices one period at a time. For text, we used pre-trained ELMO embedding. For images, we used pre-trained ResNet 50 embedding. The structure is shown in Figure ~\ref{fig:elmo_st}. This neural network takes the pre-computed ELMO text embedding (of dimension 256) and the pre-computed ResNet50 image embedding (of dimension 2048) as input, transforms through 1 to 3 fully connected hidden layers with dropout, and a final linear layer maps the last hidden layer of neurons to the one-dimensional output, which is the predicted hedonic price. We trained one model for each time period, so in total there are $T$ neural networks for $T$ time period. 

\begin{figure}
\begin{center}
\def\layersep{2.25cm}
\begin{tikzpicture}[shorten >=1pt,->,draw=black!50, node distance=\layersep]
    \tikzstyle{neuron}=[rectangle,fill=black!25,minimum size=21pt,inner sep=0pt]
    \tikzstyle{layer}=[rectangle,fill=black!25,minimum size=21pt,inner sep=0pt]

    \tikzstyle{input neuron}=[neuron, fill=green!60];
    \tikzstyle{output neuron}=[neuron, fill=magenta!60];
    \tikzstyle{hidden neuron}=[neuron, fill=blue!60];
    \tikzstyle{annot} = [text width=6em, text centered];

    \node[input neuron, pin={[pin edge={black,  <-}, blue] below:ELMO}] (I-1) at (2,-.5) {$W$};
    \node[input neuron, pin={[pin edge={black,  <-}, blue] below:ResNet50}] (I-2) at (4,-.5) {$I$};

    \foreach \name / \y in {1,...,7}  	
	\node[hidden neuron] (H-1\y) at (\y-1, 1.25) {$E_{\y}^1$};	
    \foreach \name / \y in {1,...,5}
	\node[hidden neuron] (H-2\y) at (\y cm, 2.75) {$V_{\y}$};
   
    \node[output neuron] (O) at (3, 4.5) {$H_t$};

    \foreach \source in {1,2}
        \foreach \dest in {1,...,7}
            \path (I-\source) edge (H-1\dest);
    \foreach \source in {1,...,7}
        \foreach \dest in {1,...,5}
            \path (H-1\source) edge (H-2\dest);

    \foreach \source in {1,...,5}
        \path (H-2\source) edge (O);
    
    \node[annot] (hidden1) at (-1.5, 2) {Fully Connected Layers};
    \node[annot] (input) at (-1.5,-.5) {Input Embeddings};
    \node[annot] (output) at (-1.5, 4.5) {Output};
\end{tikzpicture}

\end{center}
\caption{\small SingleTask + ELMO model. Product text is mapped to $W$ and the image is mapped to $I$ of dimensions $256$ and $2048$. For illustration purposes, we only show two hidden layers with dimensions 7 and 5 respectively. In practice, we use three layers with dimensions 2048, 1024, and 256. The output is price for one time period $t$.}\label{fig:elmo_st}
\end{figure}

\subsection{Model 2: BERT + Multitask model} 

The second neural network model that we tested is the Multitask NN with pre-trained BERT embeddings. The BERT embeddings are precomputed from a multi-lingual BERT model trained by Google. The input is ResNet50 image embeddings (of dimension 2048) and concatenated BERT sentence embeddings for the title, brand, description, and bullet points (of dimension 768 $\cdot$ 4 = 3072). The multitask NN has 1 to 3 dense layers and the output is a $T$ dimensional vector, which represents the hedonic price for each of the $T$ time periods.

\begin{figure}
\begin{center}
\def\layersep{2.25cm}
\begin{tikzpicture}[shorten >=1pt,->,draw=black!50, node distance=\layersep]
    \tikzstyle{neuron}=[rectangle,fill=black!25,minimum size=21pt,inner sep=0pt]
    \tikzstyle{layer}=[rectangle,fill=black!25,minimum size=21pt,inner sep=0pt]

    \tikzstyle{input neuron}=[neuron, fill=green!60];
    \tikzstyle{output neuron}=[neuron, fill=magenta!60];
    \tikzstyle{hidden neuron}=[neuron, fill=blue!60];
    \tikzstyle{annot} = [text width=6em, text centered];

    \node[input neuron, pin={[pin edge={black,  <-}, blue] below:BERT}] (I-1) at (2,-.5) {$W$};
    \node[input neuron, pin={[pin edge={black,  <-}, blue] below:ResNet50}] (I-2) at (4,-.5) {$I$};

    \foreach \name / \y in {1,...,7}  	
	\node[hidden neuron] (H-1\y) at (\y-1, 1.25) {$E_{\y}^1$};	
    \foreach \name / \y in {1,...,5}
	\node[hidden neuron] (H-2\y) at (\y cm, 2.75) {$V_{\y}$};
   
    \foreach \name / \y in {1,...,3}
    	\node[output neuron] (O-\y) at (\y+1, 4.5) {$H_\y$};

    \foreach \source in {1,2}
        \foreach \dest in {1,...,7}
            \path (I-\source) edge (H-1\dest);
    \foreach \source in {1,...,7}
        \foreach \dest in {1,...,5}
            \path (H-1\source) edge (H-2\dest);

    \foreach \source in {1,...,5}
    	\foreach \dest in {1,...,3}
        	    \path (H-2\source) edge (O-\dest);
    
    \node[annot] (hidden1) at (-1.5, 2) {Fully Connected Layers};
    \node[annot] (input) at (-1.5,-.5) {Input Embeddings};
    \node[annot] (output) at (-1.5, 4.5) {Output};
\end{tikzpicture}

\end{center}
\caption{\small MultiTask + BERT. Product text is mapped to $W$ and image is mapped to $I$ of dimensions $3072$ and $2048$
respectively. For illustration purpose, we only show two hidden layers with dimension 7 and 5 respectively, and output is of dimension 3. In practice, we use three layers with dimension 2048, 1024, and 256, and the output is a price vector over $T=72$ time periods.}\label{fig:BERT_mt}
\end{figure}

\subsection{Fine-Tuned BERT + Multitask model}
 
In the last experiment, we used the end-to-end training framework to fine-tune a BERT model for hedonic price prediction. The model takes raw product text as input, tokenized using the WordPiece tokenizer and truncated / padded to a maximum sequence length $d_S$, and run through a BERT base model which consists of 12 transformer blocks. Then the sequence output (of dimension $d_S \times d_T$) from the transformer blocks is aggregated through a Global Average Pooling layer to product embedding (of dimension $d_T$). Then the product embedding is linearly mapped to the output which is $T$-dimensional hedonic price vector for $T$ time periods. In our experiment, we use $d_S=512$ and $d_T=768$. 

The loss function is the same as in Model 2, which combines the weighted squared error term and a regularization term that controls volatility. The weights from all or some layers of the transformer blocks are fine-tuned for the pricing task. Figure \ref{fig:e2e_mt} is a simple illustration of the end-to-end BERT + Multitask model structure.
 
We have experimented with different numbers of fine-tuned layers. Results show that fine-tuning more layers improve model performance by a large margin, but it also takes much longer to train. This part of the work is not included in this paper,
and we are continuing to explore this research direction. 

\begin{figure}
\begin{center}
\def\layersep{2.25cm}
\begin{tikzpicture}[shorten >=1pt,->,draw=black!50, node distance=\layersep]
    \tikzstyle{neuron}=[rectangle,fill=black!25,minimum size=21pt,inner sep=0pt]
    \tikzstyle{layer}=[rectangle,fill=black!25,minimum size=21pt,inner sep=0pt]

    \tikzstyle{input neuron}=[neuron, fill=green!60];
    \tikzstyle{output neuron}=[neuron, fill=magenta!60];
    \tikzstyle{hidden neuron}=[neuron, fill=blue!60];
    \tikzstyle{annot} = [text width=8em, text centered];

    \foreach \y in {1,...,5} {
        \node[input neuron, pin={[pin edge={black,  <-}, blue] below:$t_{\y}$}] (I-\y) at (\y,-.5) {$U_{\y}$};

        \foreach \z / \zpos in {1/1,2/2.5} 
            \node[hidden neuron] (H-\z-\y) at (\y cm, \zpos) {$E_{\y\z}$};
    };

    \foreach \y in {1,...,3}
        \node[output neuron] (O-\y) at (\y +1, 4.75) {$H_{\y}$};

    \foreach \source in {1,...,5}
        \foreach \dest in {1,...,5}{
             \path (I-\source) edge (H-1-\dest);
             \path (H-1-\source) edge (H-2-\dest);
        };
        
    \foreach \source in {1,...,5}
    	\foreach \dest in {1,...,3}
        	    \path (H-2-\source) edge (O-\dest);

    \filldraw[draw=black,pattern= north east lines] (0.25,3.15) rectangle (5.75,4.15);
    
    \node[annot] (hidden1) at (-1, 1.75) {Transformer Blocks };
    \node[annot] (input) at (-1,-.5) {Context-free + Positional Embedding};
    \node[annot] (output) at (-1, 4.75) {Output};
    \node[annot] (softmax) at (-1, 3.5) {Global Average Pooling};

\end{tikzpicture}

\end{center}
\caption{\small MultiTask + Fine-tuned BERT. Product text (sentence) is tokenized and padded to $X$, where components represent the context-free input embedding plus a positional encoding for a token (word). Then the input is fed into a BERT model which consists of 12 layers of transformer blocks and outputs $T$ dimensional price vector.  For illustration purposes, we only show 5 tokens and two transformer blocks.}\label{fig:e2e_mt}
\end{figure}

\section{Additional details}
\subsection{Identification of average marginal willingness to pay via the hedonic price function} \label{appendix:identification}
We briefly sketch identification of the average derivative of a structural function under suitable assumptions. For illustration, we assume with some loss of generality that unobservable product characteristics $U$ are independent of observable characteristics $X$ at equilibrium.

This argument supports our characterization (made in Section \ref{sec:hedonic-indices}) that the average partial derivative of the hedonic price function (with respect to the equilibrium distribution of observed and unobserved characteristics) is equal to the average marginal willingness to pay for a particular characteristic.

In the hedonic equilbrium (see e.g.~\citealp{ekeland2004identification}), quantities $q_t(x,u)$ and prices $h_t(x,u)$ are well-defined as functions of observed characteristics $x \in \mathcal{X}$ and unobserved characteristics $u \in \mathcal{U}$, for $\mathcal{X}, \mathcal{U} \subset \mathbb{R}^d$. Moreover, at equilibrium, the distribution of characteristics $(X, U) \sim F_t$ is fixed. We denote by $F_t^X$ and $F_t^U$ the associated marginal distributions.

In order to make the sketch, we assume independence of $U$ and $X$ under the distribution $F_t$. We further assume that the function $h_t$ is bounded and that the functions $\{\nabla_x h_t(x,u)\}_{u \in \mathcal{U}}$ are equicontinuous. 

By independence of $X$ and $U$ under $F_t$, the average structural function
\[h_t(x) = \int h_t(x,u) dF_t^U(u)\] coincides with the hedonic regression function, in the sense that
 $h_t(X) = \mathbb{E}_{F_t}[h_t(X,U)|X]$ almost surely. To prove this claim, let $\psi$ denote any bounded, $X$-measurable test function, and note that by independence the joint distribution induced by $F_t$ coincides with that of the product measure $F_t^X \otimes F_t^U$. Thus,
 \begin{align*}
 \mathbb{E}_{F_t}[\psi(X) h_t(X,U)] &= \int h_t(x,u) \psi(x)\, dF_t(x,u)  \\
 &= \int h_t(x,u)\psi(x) \,d \{F_t^X \otimes F_t^U\}(x,u). \\
 \intertext{By Fubini's theorem, using boundedness of $\psi$ and $h_t$, this is the same as}
  &= \int \left[\int h_t(x,u)dF_t^U(u) \right] \psi(x) \, dF_t^X(x)  \\
  &= \int h_t(x) \psi(x) \,dF_t^X(u) = \mathbb{E}_{F_t}[h_t(X)\psi(X)].
 \end{align*}
 By the definition of the conditional expectation, we conclude that 
 \[h_t(X) = \mathbb{E}_{F_t}[ h_t(X,U)|X].\]
 Under standard conditions (say, if the functions $\{\nabla_x h_t(x,u)\}_{u \in \mathcal{U}}$ are equicontinuous), we may then differentiate under the integral to obtain 
 \[\frac{\partial}{\partial x_k} h_t(x) = \frac{\partial}{\partial x_k} \int  h_t(x,u) \, dF_t^U(u)  = \int \frac{\partial}{\partial x_k} h_t(x,u) \, dF_t^U(u). \]
Thus, after integrating both sides with respect to $dF_t^X$, we find that the average derivative of the hedonic price function (left-hand side) coincides with the average marginal willingness to pay for characteristics $x_k$ (right-hand side).

\subsection{Impact of product variations}
\label{subsec:siblings}
In Amazon's data, products are distinguished by their so-called ``Amazon standard identification number,'' or ASIN. These numbers are assigned at the finest-possible granularity; for example, multiple editions of the same book are assigned different ASINs, as are different colors or sizes of the same jacket. Such ``child ASINs'' are then grouped together into ``parent ASINs'' which include every size and color of a given jacket, or every edition of a particular book.

As we discussed in Section \ref{subsec:defining-products} in the main text, we define a distinct product for each ``child ASIN.'' This seems sensible for two reasons. For one, in the apparel sector, different sizes and colors can often sell for different prices. For another, this choice gives the neural network freedom to capture important price differences between similar products when data are sufficiently rich, while preserving the possibility of lumping together similar products when fewer transactions are available. Such trade-offs will naturally occur due to regularization of the network's weights. 

In principle, the low out-of-sample error we report could be driven by good performance for such ``sibling'' products that happen to be split across training and test datasets. We first remark that good performance in these cases is still helpful, since product entry and exit often occurs for some colors/sizes and not others. Fortunately, Amazon tabulates ``sibling'' products (e.g.~variations in size and color, or newer versions of existing products). We found that of the roughly 2.6 million products in the test dataset,  
 only 2.93\% (roughly 60 thousand) of those products had a relative in the training dataset, limiting the potential impact of these examples on the reported accuracy.
 
\begin{figure}
    \centering
    \includegraphics[width=0.43\linewidth]{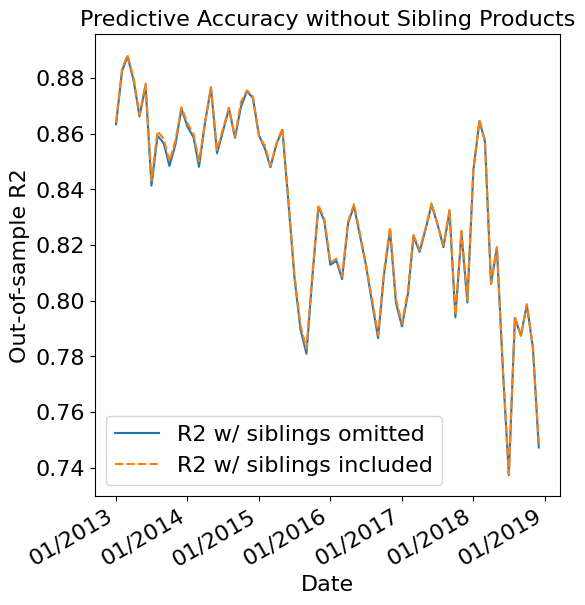}
    \includegraphics[width=0.45\linewidth]{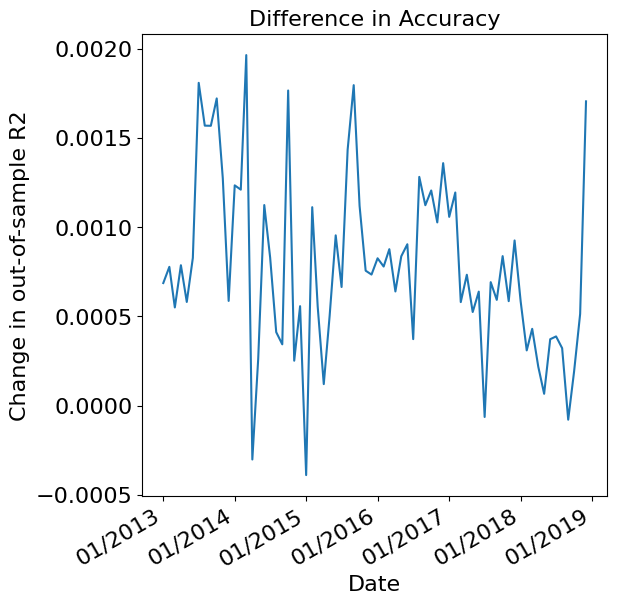}
    \caption{Predictive accuracy with and without sibling products. On the left, we have plotted the $R^2$ obtained both with (orange, dashed) and without (blue, solid) including products that have a sibling in the training dataset (e.g.,~a variation in size or color, or a newer version). The lines differ imperceptibly\textemdash the difference between the two is plotted on the right hand side; it hovers around $0.001$.}
    \label{fig:accuracy}
\end{figure}

As a final check, we plotted the difference in $R^2$ within each period when all products in the test dataset which have a relative in the training dataset are removed from consideration. We found that the $R^2$ was reduced by a negligible amount, on the order of $0.001$. This is detailed in Figure \ref{fig:accuracy}.
 
\end{document}

%% file: header/packages.tex
\usepackage[margin=1.25in]{geometry}
\usepackage{fourier}
\usepackage{amsmath, amssymb, graphicx, url, mathtools}
\usepackage[round,authoryear]{natbib}
\usepackage{xcolor}
\usepackage{wrapfig}
\usepackage{tikz}
\usepackage{float}
\usepackage{placeins}

\usepackage[normalem]{ulem}
\usepackage{subfiles}
\usepackage[multiple]{footmisc}
\usepackage[inline]{enumitem}
\usepackage{calc,tikz,nicefrac,xifthen,graphicx,subcaption,url}

\usepackage{csquotes}
\let\itenquote\enquote
\renewcommand*{\enquote}[1]{\itenquote{{\itshape#1}}}

\usetikzlibrary{patterns}
\usetikzlibrary{math}
\usetikzlibrary{shapes.misc}
\usetikzlibrary{decorations.pathreplacing}
\usetikzlibrary{patterns}

%% file: header/macros.tex
\DeclareMathOperator{\Var}{Var}

\newcommand{\Ep}{{\mathrm{E}}}

\newcommand{\mC}{\mathcal{C}}

%% file: figures/elmo-diagram.tex
\def\layersep{2.25cm}
\begin{tikzpicture}[shorten >=1pt,->,draw=black!50, node distance=\layersep]
    \tikzstyle{every pin edge}=[<-,shorten <=1pt]
    \tikzstyle{neuron}=[rectangle,fill=black!25,minimum size=21pt,inner sep=0pt]

    \tikzstyle{layer}=[rectangle,fill=black!25,minimum size=21pt,inner sep=0pt]

    \tikzstyle{input neuron}=[neuron, minimum size=21pt, fill=green!60];
    \tikzstyle{output neuron}=[neuron, minimum size=21pt, fill=magenta!60];
    \tikzstyle{hidden neuron}=[neuron, minimum size=21pt, fill=blue!60];
    \tikzstyle{annot} = [text width=6em, text centered]

    \foreach \name / \y in {1/-1.5,2/-0.5,3/.5,4/1.5}
        \node[input neuron, pin=below:$t_{\name}$] (I-\name) at (\y,-.5) {$x_{\name}$};

    \foreach \z in {1,2} {
        \foreach \y / \yy in {1/-4,2/-3,3/-2,4/-1} {
            \path[yshift=0.25cm]
                node[hidden neuron] (f-\y-\z) at (\y cm, \z) {$w^f_{\y\z}$};
            \path[yshift=0.25cm]
                node[hidden neuron] (b-\y-\z) at (\yy cm, \z) {$w^b_{\y\z}$};
        };
        \foreach \y/\yy in {1/2,2/3,3/4} {
            \path (f-\y-\z) edge (f-\yy-\z);
            \path (b-\yy-\z) edge (b-\y-\z);
        };
    };

    \foreach \y in {1,...,4} {
            \path (I-\y) edge (f-\y-1);
            \path (I-\y) edge (b-\y-1);
    };

    \foreach \z/\zz in {1/2} {
        \foreach \y in {1,...,4} {
            \path (f-\y-\z) edge (f-\y-\zz);
            \path (b-\y-\z) edge (b-\y-\zz);
        };
    };

\filldraw[draw=gray,pattern= north east lines] (-4.25,3) rectangle (4.25,4);

\foreach \y / \yy in {1/-4,2/-3,3/-2,4/-1} {
    \path node[output neuron] (of-\y) at (\y cm, 4.75) {$p^{f}_{\y}$};
    \path (f-\y-2) edge (of-\y);
    \path node[output neuron] (ob-\y) at (\yy cm, 4.75) {$p^{b}_{\y}$};
    \path (b-\y-2) edge (ob-\y);
    };
   
    \node[annot] (hidden) at (-6, 1.75) {Hidden Layers};
    \node[annot] (input) at (-6,-.5) {Context-free Embedding};
    \node[annot] (output) at (-6, 4.75) {Outputs};
    \node[annot] (softmax) at (-6, 3.5) {Softmax (Logit)};
\end{tikzpicture}

%% file: figures/bert-diagram.tex
\usetikzlibrary{positioning,fit}
\begin{tikzpicture}[shorten >=1pt,->,draw=black!50, node distance=2em]
    \tikzstyle{neuron}=[rectangle,fill=black!25,minimum size=20pt,inner sep=0pt]

    \tikzstyle{layer}=[rectangle,fill=black!25,minimum size=20pt,inner sep=0pt]

    \tikzstyle{input neuron}=[neuron, fill=green!60];
    \tikzstyle{output neuron}=[neuron, fill=magenta!60];
    \tikzstyle{hidden neuron}=[neuron, fill=blue!60];
    \tikzstyle{annot} = [text width=6em, text centered];

    \foreach \y in {1,...,5} {
        \node[input neuron, pin={[pin edge={black,  <-}, blue] below:$t_{\y}$}] (I-\y) at (\y,-.5) {$x_{\y}$};

        \foreach \z / \zpos in {1/1,2/2.5} {
            \node[hidden neuron] (h-\y-\z) at (\y cm, \zpos) {};
        };  
        
            \path (I-\y) edge (h-\y-1);
            \path (h-\y-1) edge (h-\y-2);
        \node[output neuron] (oh-\y) at (\y cm, 4.75) {$p^{b}_{\y}$};
        \path (h-\y-2) edge (oh-\y);
    };

    \foreach \z in {1,2} {
        \node[hidden neuron, fit={(h-1-\z) (h-5-\z)},text centered, text height=14pt]  {$T_\z$};
    }

    \filldraw[draw=gray!50,pattern= north east lines] (0.25,3.15) rectangle (5.75,4.15);

    \node[annot] (hidden1) at (-1.5, 1.75) {Transformers};
    \node[annot] (input) at (-1.5,-.5) {Context-free + Positional Embedding};
    \node[annot] (output) at (-1.5, 4.75) {Output};
    \node[annot] (softmax) at (-1.5, 3.5) {Feed-forward};
\end{tikzpicture}